\RequirePackage{ifpdf}
\ifpdf 
\documentclass[pdftex]{sigma}
\else
\documentclass{sigma}
\fi

\begin{document}

\renewcommand{\textfraction}{0.01}
\renewcommand{\topfraction}{0.99}

\allowdisplaybreaks

\renewcommand{\PaperNumber}{045}

\FirstPageHeading

\renewcommand{\thefootnote}{$\star$}

\ShortArticleName{Rapidities and Observable 3-Velocities in the
Flat Finslerian Event Space}

\ArticleName{Rapidities and Observable 3-Velocities\\ in the Flat
Finslerian Event Space\\ with Entirely Broken 3D
Isotropy\footnote{This paper is a contribution to the Proceedings
of the Seventh International Conference ``Symmetry in Nonlinear
Mathematical Physics'' (June 24--30, 2007, Kyiv, Ukraine). The
full collection is available at
\href{http://www.emis.de/journals/SIGMA/symmetry2007.html}{http://www.emis.de/journals/SIGMA/symmetry2007.html}}}

\Author{George Yu. BOGOSLOVSKY} \AuthorNameForHeading{G.Yu.
Bogoslovsky}

\Address{Skobeltsyn Institute of Nuclear Physics, Moscow State
University, 119991 Moscow, Russia}
\Email{\href{mailto:bogoslov@theory.sinp.msu.ru}{bogoslov@theory.sinp.msu.ru}}

\ArticleDates{Received December 09, 2007, in f\/inal form May 08,
2008; Published online May 26, 2008}

\Abstract{We study the geometric phase transitions that accompany
the dynamic rearrangement of vacuum under spontaneous violation of
initial gauge symmetry. The rearrangement may give rise to
condensates of three types, namely the scalar, axially symmetric,
and entirely anisotropic condensates. The f\/lat space-time keeps
being the Minkowski space in the only case of scalar condensate.
The anisotropic condensate having arisen, the respective
anisotropy occurs also in space-time. In this case the space-time
f\/illed with axially symmetric condensate proves to be a f\/lat
relativistically invariant Finslerian space with partially broken
3D isotropy, while the space-time f\/illed with entirely
anisotropic condensate proves to be a f\/lat relativistically
invariant Finslerian space with entirely broken 3D isotropy. The
two Finslerian space types are described brief\/ly in the extended
introduction to the work, while the original part of the latter is
devoted to determining observable 3-velocities in the entirely
anisotropic Finslerian event space. The main dif\/f\/iculties that
are overcome in solving that problem arose from the nonstandard
form of the light cone equation and from the necessity of correct
introducing of a norm in the linear vector space of rapidities.}


\Classification{53C60; 53C80; 83A05; 81T13; 81R40}

\vspace{-2mm}

\section{Introduction}

Of late years, the Lorentz symmetry violation has been widely
canvassed in  literature. The interest in the topic has arisen to
a large extent from constructing the string-motivated
phenomenological theory referred to as the Standard Model
Extension (SME)
\cite{1kostelecky&samuel,2colladay&kostelecky,3kosteleckyEd}.

Against the background of the research made in terms of SME, the
alternative (so called Finslerian) approach to the Lorentz
symmetry violation (see, in particular,
\cite{4bogoslovsky1977,5tavakol1985,6tavakol1986,7bogoslovskyBook,
8bogoslovsky199394,9bogoslovsky&goenner2004,10bogoslovsky2005,11kostelecky2004,
12bailey&kostelecky2006,13girelly&liberati&sindoni2007,14ghosh&pal2007,15bogoslovsky2007,
16gibbons&gomis&pope2007,17mavromatos2007,18sindoni2007}) gets
ever more popular. The latter is based on the Finslerian, rather
than pseudo-Euclidean, geometric model of f\/lat event space. The
great merit of the Finslerian model that predetermines its role in
developing the fundamental interaction theory and the relativistic
astrophysics arises from the fact that the model leads to Lorentz
symmetry violation without violating the relativistic symmetry.

Apart from the Minkowski event space, there exist only two types
of f\/lat Finslerian spaces that exhibit relativistic symmetry,
i.e.\ the symmetry corresponding to the Lorentz boosts. The
f\/irst type Finslerian space is a space with partially broken 3D
isotropy, while the second one exhibits an entirely broken 3D
isotropy. Described below will be their basic properties (for more
details
see~\cite{8bogoslovsky199394,19bogoslovsky&goenner199899}).

\subsection[The flat relativistically-symmetric Finslerian event space with partially broken 3D isotropy]{The f\/lat relativistically-symmetric Finslerian event space\\ with partially broken 3D isotropy}

The metric of the said space-time obtained in
\cite{20bogoslovsky1973} is
\begin{gather}\label{eq1}
ds^2=\left[\frac{(dx_0-\boldsymbol\nu d\boldsymbol
x)^2}{dx_0^2-d\boldsymbol x^{2}}\right]^r (dx_0^2-d\boldsymbol
x^{2}).
\end{gather}
This metric depends on two constant parameters, $r$ and
$\boldsymbol\nu$, and generalizes the Minkowski metric, in which
case $r$   determines the magnitude of space anisotropy, thereby
characterizing the degree of deviation of metric \eqref{eq1} from
the Minkowski metric. Instead of the 3-parameter group of
rotations of Minkowski space, the given space permits only the
1-parameter group of rotations around unit vector $\boldsymbol\nu$
that indicates a physically preferred direction in 3D space. No
changes occur for translational symmetry, and space-time
translations leave metric \eqref{eq1} invariant. As to the
transformations that relate the various inertial frames to each
other, the ordinary Lorentz boosts modify metric~\eqref{eq1}
conformally. Therefore, they do not belong to the isometry group
of space-time~\eqref{eq1}. Using them, however, we can construct
invariance transformations~\cite{4bogoslovsky1977} for
metric~\eqref{eq1}. The corresponding generalized Lorentz
transformations (generalized Lorentz boosts) prove to be
\begin{gather}\label{eq2}
x'^i=D(\boldsymbol v,\boldsymbol\nu )R^i_j(\boldsymbol
v,\boldsymbol\nu )L^j_k(\boldsymbol v) x^k,
\end{gather}
where $\boldsymbol v$ denotes the velocities of moving (primed)
inertial reference frames, the matrices $L^j_k(\boldsymbol v)$
represent the ordinary Lorentz boosts, the matrices
$R^i_j(\boldsymbol v,\boldsymbol\nu )$ represent additional
rotations of the spatial axes of the moving frames around the
vectors $[\boldsymbol v\boldsymbol\nu ]$ through the angles
\begin{gather*}
\varphi=\arccos\left\{ 1-\frac{(1-\sqrt{1-\boldsymbol v^{2}/c^2})
[\boldsymbol v\boldsymbol\nu ]^2}{ (1-\boldsymbol v\boldsymbol\nu
/c)\boldsymbol v^{2}}\right\}
\end{gather*}
of relativistic aberration of $\boldsymbol\nu,$ and the diagonal
matrices
\begin{gather*}\nonumber
D(\boldsymbol v,\boldsymbol\nu )=\left(\frac{1-\boldsymbol
v\boldsymbol\nu /c} {\sqrt{1-\boldsymbol v^{2}/c^2}} \right)^rI
\end{gather*}
stand for the additional dilatational transformations of the event
coordinates.

In contrast to the ordinary Lorentz boosts, the generalized ones
\eqref{eq2} make up a 3-parameter noncompact group with generators
$X_1$, $X_2$, $X_3$. Thus, with the inclusion of the 1-parameter
group of rotations around the preferred direction $\boldsymbol\nu$
and of the 4-parameter group of translations, the inhomogeneous
group of isometries of event space \eqref{eq1} proves to have
eight parameters. To obtain the simplest representation of its
generators, it is suf\/f\/icient to choose the third space axis
along $\boldsymbol\nu$ and, after that, to use the inf\/initesimal
form of transformations \eqref{eq2}. As a result we get
\begin{gather}
X_1=-(x^1p_0+x^0p_1)-(x^1p_3-x^3p_1),\qquad
X_2=-(x^2p_0+x^0p_2)+(x^3p_2-x^2p_3),\nonumber\\
X_3=-rx^ip_i-(x^3p_0+x^0p_3),\qquad R_3=x^2p_1-x^1p_2, \qquad
p_i=\partial /\partial x^i.\label{eq3}
\end{gather}
According to \cite{4bogoslovsky1977}, the above generators satisfy
the commutation relations
\begin{alignat}{5}
&  [X_1X_2]=0, && [R_3X_3]=0, && &&  & \nonumber\\
&  \left[X_3X_1\right]=X_1 , && [R_3X_1]=X_2 , &&  &&  &\nonumber\\
&  \left[X_3X_2\right]=X_2, && [R_3X_2]=-X_1, && && &\nonumber\\
&  \left[p_i p_j\right]=0, && &&  && &\nonumber \\
&  \left[X_1p_0\right]=p_1,&& [X_2p_0]=p_2, && [X_3p_0]=rp_0+p_3,
\qquad &&
  [R_3p_0]=0, &\label{eq4}\\
&  \left[X_1p_1\right]=p_0+p_3,\qquad && [X_2p_1]=0, &&
[X_3p_1]=rp_1, &&
  [R_3p_1]=p_2, & \nonumber\\
&  \left[X_1p_2\right]=0, && [X_2p_2]=p_0+p_3,\qquad  &&
[X_3p_2]=rp_2, &&
  [R_3p_2]=-p_1, & \nonumber \\
&  \left[X_1p_3\right]=-p_1, && [X_2p_3]=-p_2, &&
[X_3p_3]=rp_3+p_0, &&
  [R_3p_3]=0. & \nonumber
\end{alignat}
From this it is seen that the homogeneous isometry group of
axially symmetric Finslerian event space \eqref{eq1} contains four
parameters (generators $X_1$, $X_2$, $X_3$ and $R_3$). Being a
subgroup of the 11-parameter similitude (Weyl) group
\cite{21patera&winternitz&zassenhaus1975}, it is isomorphic to the
respective 4-parameter subgroup (with generators $X_1, X_2,
X_3|_{r=0}$ and $R_3$) of the homogeneous Lorentz group. Since the
6-parameter homogeneous Lorentz group has no 5-parameter subgroup,
while the 4-parameter subgroup is unique (up to isomorphisms)
\cite{22wintwrnitz&fris1965}, the transition from the Minkowski
space to the axially symmetric Finslerian space~\eqref{eq1}
implies a minimum possible symmetry-breaking of the Lorentz
symmetry, in which case the relativistic symmetry represented now
by the generalized Lorentz boosts~\eqref{eq2} remains preserved.

Since the light signals propagate in Finslerian event space
\eqref{eq1} quite in the same manner as in the Minkowski space,
the use~\cite{23bogoslovsky1982} of the Einstein procedure of
exchange of light signals  makes it possible to conclude that the
coordinates $x_0$ and $\boldsymbol x$ used to prescribe
metric~\eqref{eq1} have the meaning of observable Galilean
coordinates of events. Accordingly, the coordinate velocity
$\boldsymbol v=d\boldsymbol x/dx_0$   is an observable, so the
Einstein law of 3-velocity addition remains valid in this case.
Formally, this means that the transition from Minkowski space to
Finslerian space \eqref{eq1} fails to alter the geometric
properties of 3-velocity space; namely, the latter is again a
Lobachevski space with metric
\begin{gather*}
dl^2=\frac{(d\boldsymbol v)^2-[{\boldsymbol v}d{\boldsymbol
v}]^2}{(1-{\boldsymbol v}^2)^2},
\end{gather*}
with the orthogonal components of $\boldsymbol v$ being the
Beltrami coordinates therein.

Important information about the physical displays of space
anisotropy can be obtained by examining the Lagrange function
\cite{4bogoslovsky1977} that corresponds to Finslerian metric
\eqref{eq1}. The function describes the peculiar non-separable
interaction of a particle  with a constant (i.e.\ $r$ and
$\boldsymbol\nu$ being f\/ixed) f\/ield that characterizes space
anisotropy. As a result, we conclude that, despite the space
anisotropy, the free inertial motion of the particle remains
rectilinear and uniform as before. Contrary to the situation in
the Minkowski space, however, the particle momentum direction
fails to coincide with the particle velocity direction. In
particular, apart from its rest energy $E = mc^2,$ the particle
has the rest momentum $\boldsymbol  p=rmc\boldsymbol\nu$, while
the nonrelativistic particle inert mass (that enters Newton's
second law) proves to be  a tensor, rather than scalar
\cite{25bogoslovsky1983,24bogoslovsky1983}
\begin{gather}\label{eq6}
m_{ik}=m(1-r)({\delta}_{ik}+r{\nu}_i{\nu}_k ).
\end{gather}

As to the space anisotropy impact on the behavior of fundamental
f\/ields, the correct allowance for the impact necessitates
Finslerian generalization of the well-known f\/ield equations. In
this case, the translational invariance of Finslerian space
\eqref{eq1} makes the respective generalized f\/ield equations
admit solutions in the form of plane waves of the type $\exp
(ip_kx^k),$ where $p_k$  is the canonical 4-momentum (wave vector)
of a particle in the anisotropic space \eqref{eq1}, with $p_k$
satisfying the relativistically invariant dispersion relation
\cite{4bogoslovsky1977}
\begin{gather}\label{eq7}
\left[ {\frac{(p_0-{\boldsymbol
{p\nu}})^{2}}{{p_0}^{2}-{\boldsymbol p}^{2}}}\right]^{-r}
({p_0}^{2}-{\boldsymbol
 p}^2) = m^{2}(1-r)^{(1-r)}(1+r)^{(1+r)}.
\end{gather}
Here and henceforth we put $c = \hbar = 1$. Relation \eqref{eq7}
permits the conclusion that the anisotropy of event space
\eqref{eq1} in no way af\/fects the dynamics of massless f\/ields,
electromagnetic f\/ield in particular. Therefore, only the
fundamental massive f\/ield equations need being generalized
accordingly.

If we proceed from the Klein--Gordon equation and try its
generalization via the generalized dispersion relation \eqref{eq7}
and substitution $p_i \rightarrow i\partial/\partial x^i$, we
shall obtain a generalized Klein--Gordon equation in the form of
either linear integral equation
\cite{25bogoslovsky1983,26bogoslovsky1983} or
integro-dif\/ferential equation~\cite{16gibbons&gomis&pope2007}.
In terms of local quantum f\/ield theory, however, the Lagrangian
approach to Finslerian generali\-zation of the f\/ield equations
seems to be more adequate. The initial guideline principles of
constructing the respective generalized Lagrangians have been set
forth in~\cite{27bogoslovsky&goenner2001} and were used to
demonstrate that the real massive f\/ield does not exist as a free
f\/ield in Finslerian space-time \eqref{eq1}, but does exist as a
neutral component $\varphi_2$ of the isotopic triplet
${\boldsymbol {\varphi}}(x)= \left\{
\varphi_1(x),\varphi_2(x),-\varphi^*_1(x) \right\}$, whose
generalized Lagrangian is

\begin{gather*}
{\cal L} =
\varphi^*_{1;n}\varphi^{;n}_1+\frac{1}{2}\varphi_{2;n}\varphi^{;n}_2-\frac{m^2}{2}
(1-r^2)\left[
 \frac{{\nu}^kj_k}{(1-r)m(2\varphi^*_1\varphi_1+\varphi^2_2)}
\right]^{\frac{2r}{1+r}} (2\varphi^*_1\varphi_1 + \varphi^2_2),
\end{gather*}
where $j_k=i(\varphi^*_1\varphi_{1;k}-\varphi_1\varphi^*_{1;k}).$

As to the space anisotropy impact on the dynamics of massive
fermion f\/ield, this ef\/fect is described by the following
generalized Dirac Lagrangian \cite{9bogoslovsky&goenner2004}
\begin{gather}\label{eq9}
{\cal L}=
\frac{i}{2}\left(\bar\psi{\gamma}^{\mu}{\partial}_{\mu}\psi
-{\partial}_{\mu} \bar\psi{\gamma}^{\mu}\psi\right) -
m\left[\left( \frac{\nu_{\mu} \bar{\psi} \gamma^{\mu}
\psi}{\bar{\psi}\psi} \right)^{2} \right]^{r/2}\bar{\psi}\psi  ,
\end{gather}
 where $\nu_{\mu}=(1,-\boldsymbol\nu )$.

In contrast to the standard Dirac Lagrangian, the generalized
Dirac Lagrangian leads to nonlinear spinor equations that admit a
solution in the form of axially symmetric fermion-antifermion
condensate. The occurrence of the condensate as a physical source
of the anisotropy of f\/lat space-time~\eqref{eq1} realizes one of
the feasible mechanisms of vacuum rearrangement under spontaneous
violation of the initial gauge symmetry.

Concluding the brief description of  relativistically invariant
Finslerian space-time \eqref{eq1}, we cannot but mention the
valuable result obtained recently in the f\/ield by G.W.~Gibbons,
J.~Gomis and C.N.~Pope. We mean the CPT operator analyzed
in~\cite{16gibbons&gomis&pope2007}. It is of interest to note also
that, mostly, the above presented results were reproduced
in~\cite{16gibbons&gomis&pope2007} using the techniques of
continuous deformations of the Lie algebras and nonlinear
realizations. However, a dif\/ferent relevant notation was used
in~\cite{16gibbons&gomis&pope2007}. In particular, the parameter
that characterizes the space anisotropy magnitude was designated
$b$ instead of $r$, while the 8-parameter group of Finslerian
isometries was called DISIM${_{b}}$(2), i.e.\ Deformed
Inhomogeneous SIMilitude group that includes the 2-parameter
Abelian homogeneous noncompact subgroup. In our basis, the group
generators and Lie algebra have the form~\eqref{eq3}
and~\eqref{eq4}, respectively. As to the f\/inite transformations
that constitute the homogeneous noncompact subgroups of
DISIM${_b}$(2), they can be found in~\cite{10bogoslovsky2005}.

\subsection[The flat relativistically symmetric Finslerian event space with entirely broken 3D isotropy]{The f\/lat relativistically symmetric Finslerian event space\\ with entirely broken 3D isotropy}

The most general form of the metric of f\/lat entirely anisotropic
Finslerian event space
\begin{gather}
ds= (dx_0-dx_1-dx_2-dx_3)^{(1+r_1+r_2+r_3)/4}
(dx_0-dx_1+dx_2+dx_3)^{(1+r_1-r_2-r_3)/4}\nonumber\\
\phantom{ds=}{} \times  (dx_0+dx_1-dx_2+dx_3)^{(1-r_1+r_2-r_3)/4}
(dx_0+dx_1+dx_2-dx_3)^{(1-r_1-r_2+r_3)/4}\label{eq10}
\end{gather}
has been obtained in \cite{19bogoslovsky&goenner199899}. Three
parameters ($r_1$, $r_2$ and $r_3$) characterize the  anisotropy
of event space \eqref{eq10} and are restricted by the conditions
\begin{gather*}
1+r_1+r_2+r_3\ge 0,\qquad 1+r_1-r_2-r_3\ge 0,\\
1-r_1+r_2-r_3\ge 0,\qquad 1-r_1-r_2+r_3\ge 0.
\end{gather*}
It should be noted that, at $r_1=r_2=r_3=0,$ metric \eqref{eq10}
reduces to the fourth root of the product of four 1-forms
\begin{gather*}
ds_{_{B-M}}= [(dx_0-dx_1-dx_2-dx_3)(dx_0-dx_1+dx_2+dx_3)\\
\phantom{ds_{_{B-M}}=}{} \times
(dx_0+dx_1-dx_2+dx_3)(dx_0+dx_1+dx_2-dx_3)]^{1/4}.
\end{gather*}
In the given particular case, thus, we obtain the well-known
Berwald--Mo\'or metric \cite{28berwald1947,29moor1954} written,
however, in the basis introduced
in~\cite{19bogoslovsky&goenner199899}. As to the nonzero values of
parameters $r_i$, the values
\begin{gather*}
(r_1=1, r_2=-1, r_3=-1),\qquad (r_1=-1, r_2=-1, r_3=1),\\
(r_1=-1, r_2=1, r_3=-1), \qquad (r_1=1, r_2=1, r_3=1)
\end{gather*}
are of particular interest. The fact is that, in case the
parameters $r_i$ reach the said values, the metric \eqref{eq10},
which describes the f\/lat space-time with entirely broken 3D
isotropy, degenerates into the respective 1-forms, i.e.\ into the
total dif\/ferential of absolute time:
\begin{gather*}
ds|_{(r_1=1,r_2=-1,r_3=-1)}=dx_0-dx_1+dx_2+dx_3, \\
 ds|_{(r_1=-1,r_2=-1,r_3=1)}=dx_0+dx_1+dx_2-dx_3,\\
ds|_{(r_1=-1,r_2=1,r_3=-1)}=dx_0+dx_1-dx_2+dx_3,\\
 ds|_{(r_1=1,r_2=1,r_3=1)}=dx_0-dx_1-dx_2-dx_3.
\end{gather*}
Since the same situation arises in the case of metric~\eqref{eq1}
(the latter also degenerates at $r = 1$ into the total
dif\/ferential of absolute time), we have to
conclude~\cite{19bogoslovsky&goenner199899} that the absolute time
is not a stable degenerate state of space-time and may turn into
either partially anisotropic space-time~\eqref{eq1} or entirely
anisotropic space-time~\eqref{eq10}. In any case, the respective
geometric phase transition from the absolute time to 4D space-time
may be treated to be an Act of Creation of 3D space. This
phenomenon is accompanied by rearrangement of the vacuum state of
the system of initially massless interacting fundamental f\/ields,
resulting in that the elementary particles acquire masses. In the
case of space-time~\eqref{eq1}, the acquired particle mass is
specif\/ied by tensor~\eqref{eq6}. As to space-time~\eqref{eq10},
the acquired mass is specif\/ied by the tensor
\begin{gather}\label{eq11}
m_{ik} = m\left( \begin{array}{ccc}
(1-{r_1}^2) & (r_3-r_1r_2) & (r_2-r_1r_3) \\
(r_3-r_1r_2) & (1-{r_2}^2) & (r_1-r_2r_3) \\
(r_2-r_1r_3) & (r_1-r_2r_3) & (1-{r_3}^2)
\end{array} \right).
\end{gather}
Only after the above described process is complete, do the
concepts of spatial extension and of reference frames become
physically sensible (in a massless world, both spatial extension
of anything and one or another reference frame are meaningless to
speak of). It should be noted in this connection that, as early as
in one of the pioneer unif\/ied gauge theories (namely, the
conformal Weyl theory~\cite{30weyl191819}), the very concept of
space-time interval becomes physically meaningful only on
violating the local conformal symmetry, resulting in that the
initial massless Abelian vector gauge f\/ield acquires
mass~\cite{31bogoslovsky1992}. Finally attention should be paid to
the fact that, formally, the absolute time serves as a connecting
link, via which the correspondence principle gets satisf\/ied for
the Finslerian spaces with partially and entirely broken 3D
isotropy.

Consider now the isometry group of f\/lat Finslerian event
space~\eqref{eq10}. The homogeneous 3-parameter noncompact
isometry group, i.e.\ the relativistic symmetry group of
space-time~\eqref{eq10}, proves to be Abelian, while its
constituent transformations have the same meaning as the
conventional Lorentz boosts. The explicit form of the
transformations is
\begin{gather}\label{eq12}
x'_i=DL_{ik}x_k,
\end{gather}
where
\begin{gather*}
D=e^{-(r_1\alpha _1+r_2\alpha _2+r_3\alpha _3)} ,
\end{gather*}
the unimodular matrices $L_{ik}$ are
\begin{gather}\label{eq14}
L_{ik}=\left (
\begin{array}{rrrr}
\cal A&-\cal B&-\cal C&-\cal D\\
-\cal B&\cal A&\cal D&\cal C\\
-\cal C&\cal D&\cal A&\cal B\\
-\cal D&\cal C&\cal B&\cal A\\
\end{array}
\right ) ,
\\
{\cal A}=\cosh \alpha _1\cosh \alpha _2\cosh \alpha _3+
\sinh \alpha _1\sinh \alpha _2\sinh \alpha _3,\nonumber\\
{\cal B}=\cosh \alpha _1\sinh \alpha _2\sinh \alpha _3+
\sinh \alpha _1\cosh \alpha _2\cosh \alpha _3,\nonumber\\
{\cal C}=\cosh \alpha _1\sinh \alpha _2\cosh \alpha _3+
\sinh \alpha _1\cosh \alpha _2\sinh \alpha _3,\nonumber\\
{\cal D}=\cosh \alpha _1\cosh \alpha _2\sinh \alpha _3+ \sinh
\alpha _1\sinh \alpha _2\cosh \alpha _3,\nonumber
\end{gather}
with $\alpha _1$, $\alpha _2$, $\alpha _3$ being the group
parameters. Henceforth, the coordinate velocity components
$v_i=dx_i/dx_0$ of primed reference frame will be used as group
parameters along with the parame\-ters~$\alpha _i$. The
parame\-ters~$v_i$ and $\alpha _i$ are related to each other as
\begin{gather}\nonumber
v_1=(\tanh\alpha _1-\tanh\alpha_2\tanh\alpha _3)/(
1-\tanh\alpha_1\tanh\alpha_2\tanh\alpha_3),\\\nonumber
v_2=(\tanh\alpha _2-\tanh\alpha_1\tanh\alpha _3)/(
1-\tanh\alpha_1\tanh\alpha_2\tanh\alpha_3),\\
\label{eq15} v_3=(\tanh\alpha _3-\tanh\alpha_1\tanh\alpha _2)/(
1-\tanh\alpha_1\tanh\alpha_2\tanh\alpha_3).
\end{gather}
The inverse relations are
\begin{gather}\nonumber
\alpha _1=\frac{1}{4}\ln \frac{(1+v_1-v_2+v_3)(1+v_1+v_2-v_3)}
{(1-v_1-v_2-v_3)(1-v_1+v_2+v_3)},\\\nonumber \alpha
_2=\frac{1}{4}\ln \frac{(1-v_1+v_2+v_3)(1+v_1+v_2-v_3)}
{(1-v_1-v_2-v_3)(1+v_1-v_2+v_3)},\\\label{eq16} \alpha
_3=\frac{1}{4}\ln \frac{(1-v_1+v_2+v_3)(1+v_1-v_2+v_3)}
{(1-v_1-v_2-v_3)(1+v_1+v_2-v_3)}.
\end{gather}
As to generators $X_i$ of homogeneous 3-parameter isometry
group~\eqref{eq12} of space-time~\eqref{eq10}, they can be
presented as
\begin{gather*}\nonumber
 X_1=-r_1x_{\alpha}p_{\alpha}-(x_1p_0+
x_0p_1)+(x_2p_3+ x_3p_2),\\\nonumber
 X_2=-r_2x_{\alpha}p_{\alpha}-(x_2p_0+
x_0p_2)+(x_1p_3+
x_3p_1),\\
 X_3=-r_3x_{\alpha}p_{\alpha}-(x_3p_0+
x_0p_3)+(x_1p_2+ x_2p_1),
\end{gather*}
where $p_{\alpha}=\partial /\partial x_{\alpha}$ are generators of
the 4-parameter group of translations. Thus, on inclu\-ding the
latter, the inhomogeneous group of isometries of entirely
anisotropic Finslerian event space~\eqref{eq10} turns out to be a
7-parameter group. As to its generators, they satisfy the
commutation relations
\begin{alignat*}{4}
 & \left[X_i X_j\right]=0,\qquad && \left[p_{\alpha} p_{\beta}\right]=0, && & \nonumber\\
 & \left[X_1p_0\right]=r_1p_0+p_1,\qquad && [X_2p_0]=r_2p_0+p_2,\qquad  && [X_3p_0]=r_3p_0+p_3, &\nonumber\\
 & \left[X_1p_1\right]=r_1p_1+p_0,\qquad && [X_2p_1]=r_2p_1-p_3,\qquad && [X_3p_1]=r_3p_1-p_2, & \nonumber\\
 & \left[X_1p_2\right]=r_1p_2-p_3,\qquad & & [X_2p_2]=r_2p_2+p_0, \qquad && [X_3p_2]=r_3p_2-p_1, &\nonumber\\
 & \left[X_1p_3\right]=r_1p_3-p_2, \qquad && [X_2p_3]=r_2p_3-p_1, \qquad && [X_3p_3]=r_3p_3+p_0. & 
\end{alignat*}

Concluding the introductory part of the work, we wish to note that
the next (original) part continues studying the f\/lat
relativistically invariant Finslerian event space  with entirely
broken 3D isotropy (expression~\eqref{eq10}). The point is that,
contrary to metric~\eqref{eq1}, the coordinates $x_0$, $x_i$ used
to prescribe metric~\eqref{eq10} are not the orthogonal Galilean
coordinates of events. Accordingly, the 3-velocity $v_i=dx_i/dx_0$
is not an observable, but is only meaningful as being a coordinate
3-velocity in event space~\eqref{eq10}. It should be noted apropos
that, within the framework of conventional special relativity, the
coordinate 3-velocity that corresponds to nonorthogonal
coordinates is not an observable 3-velocity either and, to
determine the latter, there exists the well-known algorithm. From
this example alone it follows that physical identif\/ication of
various quantities and relations arising in terms of the model for
f\/lat Finslerian space of events~\eqref{eq10} deserves special
attention and turns sometimes into an independent problem. One of
the like problems that permits a novel approach to interpreting
some astrophysical observations, in particular the data relevant
to the temperature anisotropy of the microwave background
radiation, will be solved below. We mean the problem of
determining observable 3-velocities within the framework of the
model for entirely anisotropic Finslerian event
space~\eqref{eq10}. An appropriate algorithm that, in particular,
permits the magnitude of observable 3-velocity to be expressed via
the components of the latter must start being constructed by
considering the space of coordinate 3-velocities.

\section{The components of relative coordinate velocity\\ of two particles}

Obviously, the group of generalized Lorentz boosts \eqref{eq12}
that acts in entirely anisotropic event space \eqref{eq10} induces
the group of the respective transformations in the space of
coordinate 3-velocities $v_i=dx_i/dx_0.$ To obtain the
transformations, let equations~\eqref{eq12} be rewritten in terms
of coordinate dif\/ferentials
\begin{gather*}\nonumber
dx'_0=D\left({\cal A}dx_0-{\cal B}dx_1 -{\cal C}dx_2 -{\cal
D}dx_3\right),\\\nonumber dx'_1=D\left(-{\cal B}dx_0+{\cal A}dx_1
+{\cal D}dx_2 +{\cal C}dx_3\right),\\\nonumber
dx'_2=D\left(-{\cal C}dx_0+{\cal D}dx_1 +{\cal A}dx_2 +{\cal B}dx_3\right),\\
dx'_3=D\left(-{\cal D}dx_0+{\cal C}dx_1 +{\cal B}dx_2 +{\cal
A}dx_3\right).
\end{gather*}
After dividing, then, the second, third, and fourth relations by
the f\/irst one and using the fact that $v_i=dx_i/dx_0$     are
components of the coordinate 3-velocity, we get
\begin{gather}\nonumber
v'_1=\frac{-{\cal B}+{\cal A}v_1+{\cal D}v_2+{\cal C}v_3}{{\cal
A}-{\cal B}v_1-{\cal C}v_2-{\cal D}v_3},\qquad
v'_2=\frac{-{\cal C}+{\cal D}v_1+{\cal A}v_2+{\cal B}v_3}{{\cal A}-{\cal B}v_1-{\cal C}v_2-{\cal D}v_3},\\
\label{eq20} v'_3=\frac{-{\cal D}+{\cal C}v_1+{\cal B}v_2+{\cal
A}v_3}{{\cal A}-{\cal B}v_1-{\cal C}v_2-{\cal D}v_3}.
\end{gather}
The relations \eqref{eq20} interrelate the components $v_i$  and
${v'}_i$   of the coordinate 3-velocity of a particle in the
initial and primed inertial frames, respectively, with the
dependence of the involved matrix elements ${\cal A}$, ${\cal B}$,
${\cal C}$, ${\cal D}$ on the coordinate 3-velocity of the primed
reference frame being determined by relations~\eqref{eq14}
and~\eqref{eq16}. Formally, the transformations~\eqref{eq20} are a
nonlinear representation of linear group~\eqref{eq12}. Besides, it
should be noted that the representation proves to be independent
of parameters~$r_i$ and, therefore, is equally valid for the
Berwald--Mo\'or metric.

Consider now two particles. Let ${\stackrel{(2)}{v}}_{\!\!i}$ be
the coordinate 3-velocity of the second particle in the initial
reference frame, and ${\stackrel{(1)}{v}}_{\!\!i}$ be the
coordinate 3-velocity of the f\/irst particle in the same frame.
In addition, let ${\stackrel{(1)}{v}}_{\!\!i}$ be identif\/ied as
the coordinate 3-velocity of primed reference frame. As a~result,
the primed frame acquires the meaning of the rest frame of the
f\/irst particle. Finally, let the 3-velocity components of the
second particle in the rest frame of the f\/irst particle (or, in
other words, the relative velocity of the second particle with
respect to the f\/irst particle) be designated ${\stackrel{(1\to
2)}{v}}_{\!\!\!i}.$ Using these notations together with
formulas~\eqref{eq14} and \eqref{eq16}, we obtain via \eqref{eq20}
\begin{gather}\nonumber
{\stackrel{(1\to 2)}{v}}_{\!\!\!1}=\frac{-\stackrel{(1)}{\cal
B}+\stackrel{(1)}{\cal
A}{\stackrel{(2)}{v}}_{\!\!1}+\stackrel{(1)}{\cal
D}{\stackrel{(2)}{v}}_{\!\!2}+\stackrel{(1)}{\cal
C}{\stackrel{(2)}{v}}_{\!\!3}}{\stackrel{(1)}{\cal
A}-\stackrel{(1)}{\cal
B}{\stackrel{(2)}{v}}_{\!\!1}-\stackrel{(1)}{\cal
C}{\stackrel{(2)}{v}}_{\!\!2}-\stackrel{(1)}{\cal
D}{\stackrel{(2)}{v}}_{\!\!3}}, \qquad {\stackrel{(1\to
2)}{v}}_{\!\!\!2}=\frac{-\stackrel{(1)}{\cal
C}+\stackrel{(1)}{\cal
D}{\stackrel{(2)}{v}}_{\!\!1}+\stackrel{(1)}{\cal
A}{\stackrel{(2)}{v}}_{\!\!2}+\stackrel{(1)}{\cal
B}{\stackrel{(2)}{v}}_{\!\!3}}{\stackrel{(1)}{\cal
A}-\stackrel{(1)}{\cal
B}{\stackrel{(2)}{v}}_{\!\!1}-\stackrel{(1)}{\cal
C}{\stackrel{(2)}{v}}_{\!\!2}-\stackrel{(1)}{\cal
D}{\stackrel{(2)}{v}}_{\!\!3}},
\\
\label{eq21} {\stackrel{(1\to
2)}{v}}_{\!\!\!3}=\frac{-\stackrel{(1)}{\cal
D}+\stackrel{(1)}{\cal
C}{\stackrel{(2)}{v}}_{\!\!1}+\stackrel{(1)}{\cal
B}{\stackrel{(2)}{v}}_{\!\!2}+\stackrel{(1)}{\cal
A}{\stackrel{(2)}{v}}_{\!\!3}}{\stackrel{(1)}{\cal
A}-\stackrel{(1)}{\cal
B}{\stackrel{(2)}{v}}_{\!\!1}-\stackrel{(1)}{\cal
C}{\stackrel{(2)}{v}}_{\!\!2}-\stackrel{(1)}{\cal
D}{\stackrel{(2)}{v}}_{\!\!3}},
\end{gather}
where
\begin{gather*}\nonumber
\stackrel{(1)}{\cal
A}=\frac{1-\!\!\phantom{[}{\stackrel{(1)}{v}}_{\!\!1}\!\!\phantom{]}^2-\!\!\phantom{[}{\stackrel{(1)}{v}}_{\!\!2}\!\!\phantom{]}^2-\!\!\phantom{[}{\stackrel{(1)}{v}}_{\!\!3}\!\!\phantom{]}^2-2{\stackrel{(1)}{v}}_{\!\!1}{\stackrel{(1)}{v}}_{\!\!2}{\stackrel{(1)}{v}}_{\!\!3}}
{\left [\Bigl
(1\!-\!{\stackrel{(1)}{v}}_{\!\!1}\!-\!{\stackrel{(1)}{v}}_{\!\!2}\!-\!{\stackrel{(1)}{v}}_{\!\!3}\Bigr
)\Bigl
(1\!-\!{\stackrel{(1)}{v}}_{\!\!1}\!+\!{\stackrel{(1)}{v}}_{\!\!2}\!+\!{\stackrel{(1)}{v}}_{\!\!3}\Bigr
)\Bigl
(1\!+\!{\stackrel{(1)}{v}}_{\!\!1}\!-\!{\stackrel{(1)}{v}}_{\!\!2}\!+\!{\stackrel{(1)}{v}}_{\!\!3}\Bigr
)\Bigl
(1\!+\!{\stackrel{(1)}{v}}_{\!\!1}\!+\!{\stackrel{(1)}{v}}_{\!\!2}\!-\!{\stackrel{(1)}{v}}_{\!\!3}\Bigr
)\right ]^{3/4}},
\\
\stackrel{(1)}{\cal
B}=\frac{{\stackrel{(1)}{v}}_{\!\!1}+2{\stackrel{(1)}{v}}_{\!\!2}{\stackrel{(1)}{v}}_{\!\!3}-{\stackrel{(1)}{v}}_{\!\!1}\Bigl
(\!\phantom{[}{\stackrel{(1)}{v}}_{\!\!1}\!\!\phantom{]}^2-\!\!\phantom{[}{\stackrel{(1)}{v}}_{\!\!2}\!\!\phantom{]}^2-\!\!\phantom{[}{\stackrel{(1)}{v}}_{\!\!3}\!\!\phantom{]}^2\Bigr
)} {\left [\Bigl
(1\!-\!{\stackrel{(1)}{v}}_{\!\!1}\!-\!{\stackrel{(1)}{v}}_{\!\!2}\!-\!{\stackrel{(1)}{v}}_{\!\!3}\Bigr
)\Bigl
(1\!-\!{\stackrel{(1)}{v}}_{\!\!1}\!+\!{\stackrel{(1)}{v}}_{\!\!2}\!+\!{\stackrel{(1)}{v}}_{\!\!3}\Bigr
)\Bigl
(1\!+\!{\stackrel{(1)}{v}}_{\!\!1}\!-\!{\stackrel{(1)}{v}}_{\!\!2}\!+\!{\stackrel{(1)}{v}}_{\!\!3}\Bigr
)\Bigl
(1\!+\!{\stackrel{(1)}{v}}_{\!\!1}\!+\!{\stackrel{(1)}{v}}_{\!\!2}\!-\!{\stackrel{(1)}{v}}_{\!\!3}\Bigr
)\right ]^{3/4}},
\\
\stackrel{(1)}{\cal
C}=\frac{{\stackrel{(1)}{v}}_{\!\!2}+2{\stackrel{(1)}{v}}_{\!\!1}{\stackrel{(1)}{v}}_{\!\!3}-{\stackrel{(1)}{v}}_{\!\!2}\Bigl
(\!\phantom{[}{\stackrel{(1)}{v}}_{\!\!2}\!\!\phantom{]}^2-\!\!\phantom{[}{\stackrel{(1)}{v}}_{\!\!1}\!\!\phantom{]}^2-\!\!\phantom{[}{\stackrel{(1)}{v}}_{\!\!3}\!\!\phantom{]}^2\Bigr
)} {\left [\Bigl
(1\!-\!{\stackrel{(1)}{v}}_{\!\!1}\!-\!{\stackrel{(1)}{v}}_{\!\!2}\!-\!{\stackrel{(1)}{v}}_{\!\!3}\Bigr
)\Bigl
(1\!-\!{\stackrel{(1)}{v}}_{\!\!1}\!+\!{\stackrel{(1)}{v}}_{\!\!2}\!+\!{\stackrel{(1)}{v}}_{\!\!3}\Bigr
)\Bigl
(1\!+\!{\stackrel{(1)}{v}}_{\!\!1}\!-\!{\stackrel{(1)}{v}}_{\!\!2}\!+\!{\stackrel{(1)}{v}}_{\!\!3}\Bigr
)\Bigl
(1\!+\!{\stackrel{(1)}{v}}_{\!\!1}\!+\!{\stackrel{(1)}{v}}_{\!\!2}\!-\!{\stackrel{(1)}{v}}_{\!\!3}\Bigr
)\right ]^{3/4}},
\\
\stackrel{(1)}{\cal
D}=\frac{{\stackrel{(1)}{v}}_{\!\!3}+2{\stackrel{(1)}{v}}_{\!\!1}{\stackrel{(1)}{v}}_{\!\!2}-{\stackrel{(1)}{v}}_{\!\!3}\Bigl
(\!\phantom{[}{\stackrel{(1)}{v}}_{\!\!3}\!\!\phantom{]}^2-\!\!\phantom{[}{\stackrel{(1)}{v}}_{\!\!1}\!\!\phantom{]}^2-\!\!\phantom{[}{\stackrel{(1)}{v}}_{\!\!2}\!\!\phantom{]}^2\Bigr
)} {\left [\Bigl
(1\!-\!{\stackrel{(1)}{v}}_{\!\!1}\!-\!{\stackrel{(1)}{v}}_{\!\!2}\!-\!{\stackrel{(1)}{v}}_{\!\!3}\Bigr
)\Bigl
(1\!-\!{\stackrel{(1)}{v}}_{\!\!1}\!+\!{\stackrel{(1)}{v}}_{\!\!2}\!+\!{\stackrel{(1)}{v}}_{\!\!3}\Bigr
)\Bigl
(1\!+\!{\stackrel{(1)}{v}}_{\!\!1}\!-\!{\stackrel{(1)}{v}}_{\!\!2}\!+\!{\stackrel{(1)}{v}}_{\!\!3}\Bigr
)\Bigl
(1\!+\!{\stackrel{(1)}{v}}_{\!\!1}\!+\!{\stackrel{(1)}{v}}_{\!\!2}\!-\!{\stackrel{(1)}{v}}_{\!\!3}\Bigr
)\right ]^{3/4}}.
\end{gather*}
The above cumbersome formulas express the components of the
relative coordinate 3-velocity of two particles via the coordinate
3-velocity components of either particle. From the fact that the
above relations are a direct consequence of transformations
\eqref{eq12} that constitute the Abelian group with parameters
${\alpha}_i$, while ${\alpha}_i$ proper can be treated to be
rapidity components related to~$v_i$    via~\eqref{eq15}
and~\eqref{eq16}, the conclusion inevitably comes to mind that, in
terms of ${\alpha}_i$, relations~\eqref{eq21} must get
simplif\/ied signif\/icantly and take on the form
\begin{gather}\label{eq22}
{\stackrel{(1\to
2)}{\alpha}}_{\!\!i}={\stackrel{(2)}{\alpha}}_{\!i}-{\stackrel{(1)}{\alpha}}_{\!i}.
\end{gather}
The fact that the conclusion is really true can be proved as
follows. First, proceeding from \eqref{eq21}, and via direct
calculations we get the following three equations
\begin{gather*}\nonumber
\frac{\Bigl (1+{\stackrel{(1\to 2)}{v}}_{\!\!\!1}-{\stackrel{(1\to
2)}{v}}_{\!\!\!2}+{\stackrel{(1\to 2)}{v}}_{\!\!\!3}\Bigr )\Bigl
(1+{\stackrel{(1\to 2)}{v}}_{\!\!\!1}+{\stackrel{(1\to
2)}{v}}_{\!\!\!2}-{\stackrel{(1\to 2)}{v}}_{\!\!\!3}\Bigr )}
{\Bigl (1-{\stackrel{(1\to 2)}{v}}_{\!\!\!1}-{\stackrel{(1\to
2)}{v}}_{\!\!\!2}-{\stackrel{(1\to 2)}{v}}_{\!\!\!3}\Bigr )\Bigl
(1-{\stackrel{(1\to 2)}{v}}_{\!\!\!1}+{\stackrel{(1\to
2)}{v}}_{\!\!\!2}+{\stackrel{(1\to 2)}{v}}_{\!\!\!3}\Bigr )}
\\
\qquad{}=\frac{\Bigl
(1\!-\!{\stackrel{(1)}{v}}_{\!\!1}\!-\!{\stackrel{(1)}{v}}_{\!\!2}\!-\!{\stackrel{(1)}{v}}_{\!\!3}\Bigr
)\Bigl
(1\!-\!{\stackrel{(1)}{v}}_{\!\!1}\!+\!{\stackrel{(1)}{v}}_{\!\!2}\!+\!{\stackrel{(1)}{v}}_{\!\!3}\Bigr
)\Bigl
(1\!+\!{\stackrel{(2)}{v}}_{\!\!1}\!-\!{\stackrel{(2)}{v}}_{\!\!2}\!+\!{\stackrel{(2)}{v}}_{\!\!3}\Bigr
)\Bigl
(1\!+\!{\stackrel{(2)}{v}}_{\!\!1}\!+\!{\stackrel{(2)}{v}}_{\!\!2}\!-\!{\stackrel{(2)}{v}}_{\!\!3}\Bigr
)} {\Bigl
(1\!+\!{\stackrel{(1)}{v}}_{\!\!1}\!-\!{\stackrel{(1)}{v}}_{\!\!2}\!+\!{\stackrel{(1)}{v}}_{\!\!3}\Bigr
)\Bigl
(1\!+\!{\stackrel{(1)}{v}}_{\!\!1}\!+\!{\stackrel{(1)}{v}}_{\!\!2}\!-\!{\stackrel{(1)}{v}}_{\!\!3}\Bigr
)\Bigl
(1\!-\!{\stackrel{(2)}{v}}_{\!\!1}\!-\!{\stackrel{(2)}{v}}_{\!\!2}\!-\!{\stackrel{(2)}{v}}_{\!\!3}\Bigr
)\Bigl
(1\!-\!{\stackrel{(2)}{v}}_{\!\!1}\!+\!{\stackrel{(2)}{v}}_{\!\!2}\!+\!{\stackrel{(2)}{v}}_{\!\!3}\Bigr
)},
\\
\frac{\Bigl (1-{\stackrel{(1\to 2)}{v}}_{\!\!\!1}+{\stackrel{(1\to
2)}{v}}_{\!\!\!2}+{\stackrel{(1\to 2)}{v}}_{\!\!\!3}\Bigr )\Bigl
(1+{\stackrel{(1\to 2)}{v}}_{\!\!\!1}+{\stackrel{(1\to
2)}{v}}_{\!\!\!2}-{\stackrel{(1\to 2)}{v}}_{\!\!\!3}\Bigr )}
{\Bigl (1-{\stackrel{(1\to 2)}{v}}_{\!\!\!1}-{\stackrel{(1\to
2)}{v}}_{\!\!\!2}-{\stackrel{(1\to 2)}{v}}_{\!\!\!3}\Bigr )\Bigl
(1+{\stackrel{(1\to 2)}{v}}_{\!\!\!1}-{\stackrel{(1\to
2)}{v}}_{\!\!\!2}+{\stackrel{(1\to 2)}{v}}_{\!\!\!3}\Bigr )}
\\
\qquad{}=\frac{\Bigl
(1\!-\!{\stackrel{(1)}{v}}_{\!\!1}\!-\!{\stackrel{(1)}{v}}_{\!\!2}\!-\!{\stackrel{(1)}{v}}_{\!\!3}\Bigr
)\Bigl
(1\!+\!{\stackrel{(1)}{v}}_{\!\!1}\!-\!{\stackrel{(1)}{v}}_{\!\!2}\!+\!{\stackrel{(1)}{v}}_{\!\!3}\Bigr
)\Bigl
(1\!-\!{\stackrel{(2)}{v}}_{\!\!1}\!+\!{\stackrel{(2)}{v}}_{\!\!2}\!+\!{\stackrel{(2)}{v}}_{\!\!3}\Bigr
)\Bigl
(1\!+\!{\stackrel{(2)}{v}}_{\!\!1}\!+\!{\stackrel{(2)}{v}}_{\!\!2}\!-\!{\stackrel{(2)}{v}}_{\!\!3}\Bigr
)} {\Bigl
(1\!-\!{\stackrel{(1)}{v}}_{\!\!1}\!+\!{\stackrel{(1)}{v}}_{\!\!2}\!+\!{\stackrel{(1)}{v}}_{\!\!3}\Bigr
)\Bigl
(1\!+\!{\stackrel{(1)}{v}}_{\!\!1}\!+\!{\stackrel{(1)}{v}}_{\!\!2}\!-\!{\stackrel{(1)}{v}}_{\!\!3}\Bigr
)\Bigl
(1\!-\!{\stackrel{(2)}{v}}_{\!\!1}\!-\!{\stackrel{(2)}{v}}_{\!\!2}\!-\!{\stackrel{(2)}{v}}_{\!\!3}\Bigr
)\Bigl
(1\!+\!{\stackrel{(2)}{v}}_{\!\!1}\!-\!{\stackrel{(2)}{v}}_{\!\!2}\!+\!{\stackrel{(2)}{v}}_{\!\!3}\Bigr
)},
\\
\frac{\Bigl (1-{\stackrel{(1\to 2)}{v}}_{\!\!\!1}+{\stackrel{(1\to
2)}{v}}_{\!\!\!2}+{\stackrel{(1\to 2)}{v}}_{\!\!\!3}\Bigr )\Bigl
(1+{\stackrel{(1\to 2)}{v}}_{\!\!\!1}-{\stackrel{(1\to
2)}{v}}_{\!\!\!2}+{\stackrel{(1\to 2)}{v}}_{\!\!\!3}\Bigr )}
{\Bigl (1-{\stackrel{(1\to 2)}{v}}_{\!\!\!1}-{\stackrel{(1\to
2)}{v}}_{\!\!\!2}-{\stackrel{(1\to 2)}{v}}_{\!\!\!3}\Bigr )\Bigl
(1+{\stackrel{(1\to 2)}{v}}_{\!\!\!1}+{\stackrel{(1\to
2)}{v}}_{\!\!\!2}-{\stackrel{(1\to 2)}{v}}_{\!\!\!3}\Bigr )}
\\
\qquad{}=\frac{\Bigl
(1\!-\!{\stackrel{(1)}{v}}_{\!\!1}\!-\!{\stackrel{(1)}{v}}_{\!\!2}\!-\!{\stackrel{(1)}{v}}_{\!\!3}\Bigr
)\Bigl
(1\!+\!{\stackrel{(1)}{v}}_{\!\!1}\!+\!{\stackrel{(1)}{v}}_{\!\!2}\!-\!{\stackrel{(1)}{v}}_{\!\!3}\Bigr
)\Bigl
(1\!-\!{\stackrel{(2)}{v}}_{\!\!1}\!+\!{\stackrel{(2)}{v}}_{\!\!2}\!+\!{\stackrel{(2)}{v}}_{\!\!3}\Bigr
)\Bigl
(1\!+\!{\stackrel{(2)}{v}}_{\!\!1}\!-\!{\stackrel{(2)}{v}}_{\!\!2}\!+\!{\stackrel{(2)}{v}}_{\!\!3}\Bigr
)} {\Bigl
(1\!-\!{\stackrel{(1)}{v}}_{\!\!1}\!+\!{\stackrel{(1)}{v}}_{\!\!2}\!+\!{\stackrel{(1)}{v}}_{\!\!3}\Bigr
)\Bigl
(1\!+\!{\stackrel{(1)}{v}}_{\!\!1}\!-\!{\stackrel{(1)}{v}}_{\!\!2}\!+\!{\stackrel{(1)}{v}}_{\!\!3}\Bigr
)\Bigl
(1\!-\!{\stackrel{(2)}{v}}_{\!\!1}\!-\!{\stackrel{(2)}{v}}_{\!\!2}\!-\!{\stackrel{(2)}{v}}_{\!\!3}\Bigr
)\Bigl
(1\!+\!{\stackrel{(2)}{v}}_{\!\!1}\!+\!{\stackrel{(2)}{v}}_{\!\!2}\!-\!{\stackrel{(2)}{v}}_{\!\!3}\Bigr
)},
\end{gather*}
whereupon we are only to f\/ind natural logarithm of each of the
equations and use formulas \eqref{eq16}. The result thereof is
just relation~\eqref{eq22}.

According to \eqref{eq22}, the rapidities ${\alpha}_i$ form the
linear vector space, whence the problem of introducing of a norm
$\vert{\alpha}_i\vert$ in that vector space arises for us to
solve.

\section{Introducing of a norm in the vector space of rapidities}

To f\/ind $\vert{\alpha}_i\vert$ as an explicit function of
${\alpha}_i,$ attention should be paid to relation
\begin{gather}\label{eq23}
{\stackrel{(1\to 2)}{\alpha}}_{\!\!i}=-{\stackrel{(2\to
1)}{\alpha}}_{\!\!i},
\end{gather}
which means that the rapidity components of the second particle
relative to the f\/irst particle dif\/fer in sign from the
rapidity components of the f\/irst particle relative to the
second. This fact is a trivial consequence of the Abelian
structure of relativistic symmetry group \eqref{eq12} and prompts
us to write $\vert{\alpha}_i\vert
=\sqrt{{{\alpha}_1}^2+{{\alpha}_2}^2+{{\alpha}_3}^2}.$ However,
such introducing of a norm in the vector space of rapidities would
be physically incorrect because it would never lead to any
reasonable relation between the components of observable
3-velocity and its magnitude. Therefore, we will be acting as
follows.

First, proceeding from physical considerations, we shall express
the square of observable 3-velocity $V^2$ to be an explicit
function of the components of coordinate 3-velocity~$v_i$. After
that, formulas~\eqref{eq15} are to be used to represent $V^2$ as
an explicit function of the rapidity components~$\alpha_i$.
Thereby, we shall actually f\/inish the procedure of introducing
of the norm since (by def\/inition) the magnitude $V$ of
observable 3-velocity and the rapidity magnitude
$\vert{\alpha}_i\vert$ are related to one another~as
\begin{gather}\label{eq24}
V^2={\tanh}^2\vert{\alpha}_i\vert.
\end{gather}

Thus, let us, f\/irst of all, f\/ind the explicit form of function
$V^2=V^2(v_i)$. In accordance with its physical meaning, that
function, just as $V(v_i),$ must satisfy the condition
\begin{gather}\label{eq25}
V^2({\stackrel{(1\to 2)}{v}}_{\!\!i})=V^2({\stackrel{(2\to
1)}{v}}_{\!\!i}),
\end{gather}
where ${\stackrel{(1\to 2)}{v}}_{\!\!i}$ are  the components of
coordinate 3-velocity of the second particle with respect to the
f\/irst one, and ${\stackrel{(2\to 1)}{v}}_{\!\!i}$ are  the
components of coordinate 3-velocity of the f\/irst particle with
respect to the second one. Considering $v_i$ as the group
parameters that are alternative to the parameters ${\alpha}_i,$ we
can see that ${\stackrel{(1\to 2)}{v}}_{\!\!i}$ and
${\stackrel{(2\to 1)}{v}}_{\!\!i}$ are the mutually inverse
elements $g$ and $g^{-1}$ of group \eqref{eq12}.  In this case, in
terms of ${\alpha}_i,$ the transition $g\to g^{-1}$ corresponds
(see formula \eqref{eq23}) to transformation
${\alpha}_i\to{\tilde\alpha}_i=-{\alpha}_i,$ whereas in terms of
$v_i,$ the same transition corresponds (in virtue of \eqref{eq15}
and \eqref{eq16}) to transformation
\begin{gather}\nonumber
v_1\to{\tilde
v}_1=-\frac{v_1(1-v_1^2+v_2^2+v_3^2)+2v_2v_3}{1-v_1^2-v_2^2-v_3^2-2v_1v_2v_3},
\qquad v_2\to{\tilde
v}_2=-\frac{v_2(1+v_1^2-v_2^2+v_3^2)+2v_1v_3}{1-v_1^2-v_2^2-v_3^2-2v_1v_2v_3},
\\
\label{eq26} v_3\to{\tilde
v}_3=-\frac{v_3(1+v_1^2+v_2^2-v_3^2)+2v_1v_2}{1-v_1^2-v_2^2-v_3^2-2v_1v_2v_3}.
\end{gather}
Accordingly, the inverse transformations appear as
$-{\alpha}_i={\tilde\alpha}_i\to{\alpha}_i$ and
\begin{gather}\nonumber
{\tilde v}_1\to v_1=-\frac{{\tilde v}_1(1-{\tilde v}_1^2+{\tilde
v}_2^2+{\tilde v}_3^2)+2{\tilde v}_2{\tilde v}_3}{1-{\tilde
v}_1^2-{\tilde v}_2^2-{\tilde v}_3^2-2{\tilde v}_1{\tilde
v}_2{\tilde v}_3}, \qquad {\tilde v}_2\to v_2=-\frac{{\tilde
v}_2(1+{\tilde v}_1^2-{\tilde v}_2^2+{\tilde v}_3^2)+2{\tilde
v}_1{\tilde v}_3}{1-{\tilde v}_1^2-{\tilde v}_2^2-{\tilde
v}_3^2-2{\tilde v}_1{\tilde v}_2{\tilde v}_3},
\\
\label{eq27} {\tilde v}_3\to v_3=-\frac{{\tilde v}_3(1+{\tilde
v}_1^2+{\tilde v}_2^2-{\tilde v}_3^2)+2{\tilde v}_1{\tilde
v}_2}{1-{\tilde v}_1^2-{\tilde v}_2^2-{\tilde v}_3^2-2{\tilde
v}_1{\tilde v}_2{\tilde v}_3}.
\end{gather}
Now we see that equation~\eqref{eq25}, i.e.\ the f\/irst, and most
important, restriction on the sought function $V^2,$ can be
rewritten as
\begin{gather}\label{eq28}
V^2(v_i)=V^2({\tilde v}_i).
\end{gather}
Thus, the function $V^2(v_i)$ must be an invariant of
transformations \eqref{eq26} and \eqref{eq27}.

 Since, when examining a classical particle motion, we deal with causally-related events, the physically permissible range of the values of squared observable 3-velocity is restricted by the condition $0\leq V^2(v_i)\leq 1$, where
\begin{gather}\label{eq29}
V^2(0,0,0)=0
\end{gather}
in accordance with the def\/inition of coordinate 3-velocity
$v_i$. As to the range of physically permissible $v_i$ values,
that range will be shown in the next section to be restricted by
the regular tetrahedron surface presented in Fig.~\ref{fig6}. On
the tetrahedron surface proper, the coordinate 3-velocities, which
coincide with all the possible coordinate 3-velocities of a
photon, satisfy the relation
\begin{gather}\label{eq30}
(ds_{_{B-M}}/dx_0)^4=(1\!-\!v_1\!-\!v_2\!-\!v_3)(1\!-\!v_1\!+\!v_2\!+\!v_3)(1\!+\!v_1\!-\!v_2\!+\!v_3)
(1\!+\!v_1\!+\!v_2\!-\!v_3)=0.
\end{gather}
Since, on the other hand, $V^2=1$ at any direction of observable
3-velocity of the photon, relation~\eqref{eq30} implies that the
sought function $V^2(v_i)$ must satisfy the condition
\begin{gather}\label{eq31}
V^2{\bigl\vert}_{(ds_{_{_{B-M}}}/dx_{_0}=0)}=1.
\end{gather}

Apart from the stated conditions \eqref{eq28}, \eqref{eq29}, and
\eqref{eq31}, attention should be paid to the fact that, in the 2D
case, where (for instance) $dx_0\ne 0$, $dx_1\ne 0$,
$dx_2=dx_3=0,$ the Berwald--Mo\'or space coincides with the
Minkowski space, i.e.\ $ds^2_{_{B-M}}=dx^2_0-dx^2_1.$ In this
case, accordingly, the coordinate velocity $v_1$ coincides with
the observable velocity $V.$ As a result, we conclude that the
sought function $V^2(v_i)$ must also satisfy the relations
\begin{gather}\label{eq32}
V^2(v_1,0,0)=v^2_1,\qquad V^2(0,v_2,0)=v^2_2,\qquad
V^2(0,0,v_3)=v^2_3.
\end{gather}

To f\/ind the function $V^2(v_i)$ that satisf\/ies the conditions
\eqref{eq28}, \eqref{eq29}, \eqref{eq31}, and \eqref{eq32}, an
attempt will be made f\/irstly to construct some auxiliary
function $f(v_i)$ that would remain invariant under
transformations \eqref{eq26} and \eqref{eq27}. To that end,
consider two characteristic functions that enter relations
\eqref{eq26} and \eqref{eq30} above, namely
\begin{gather}\label{eq33}
f_1(v_i)=1-v_1^2-v_2^2-v_3^2-2v_1v_2v_3
\end{gather}
and
\begin{gather}\label{eq34}
f_2(v_i)=(1-v_1-v_2-v_3)(1-v_1+v_2+v_3)(1+v_1-v_2+v_3)(1+v_1+v_2-v_3).
\end{gather}
Using substitution \eqref{eq26}, we can readily prove that
$f_1({\tilde v}_i)=f^2_2(v_i)/f^3_1(v_i).$ Similarly, we can
verify that $f_2({\tilde v}_i)=f^3_2(v_i)/f^4_1(v_i).$ Therefore,
on determining the function $f(v_i)$ via relation
$f(v_i)=f_2(v_i)/f_1(v_i),$ we come to equality $f({\tilde
v}_i)=f(v_i),$ which means that the function $f(v_i)$ introduced
as above is actually an invariant of transformations \eqref{eq26}
and \eqref{eq27}. From this it follows clearly that the sought
function $V^2(v_i),$ which satisf\/ies the conditions
\eqref{eq28}, \eqref{eq29}, \eqref{eq31}, and \eqref{eq32}, is
prescribed by the relation $V^2(v_i)=1-f(v_i)=1-f_2(v_i)/f_1(v_i)$
and (owing to \eqref{eq33} and \eqref{eq34}) proves to be
\begin{gather}\label{eq35}
V^2(v_i)=1-\frac{(1 - v_1 - v_2 - v_3)(1 - v_1 + v_2 + v_3)(1 +
v_1 - v_2 + v_3)(1 + v_1 + v_2 - v_3)}{1 - v_1^2 - v_2^2 - v_3^2 -
2v_1v_2v_3}.\!\!\!\!
\end{gather}

\begin{figure}[t]
\centerline{\includegraphics[width=7.3cm]{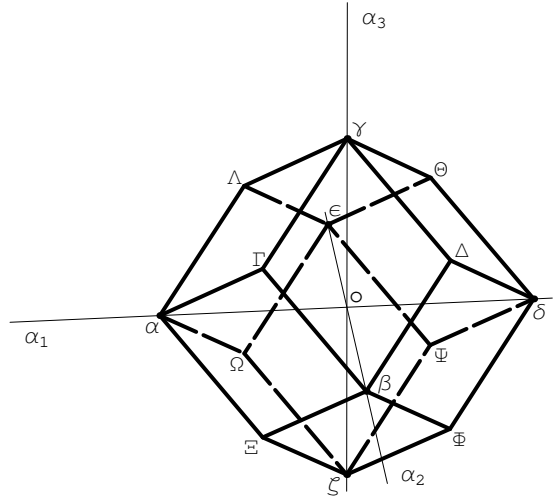}}
\vspace{-3mm} \caption{The regular rhombic dodecahedron in the
space of rapidities $\alpha _i$  as an image (at
$\tanh{\vert{\alpha _i}\vert}=1$)  of the light front
surface.}\label{fig1}
\end{figure}

Now, representing (via \eqref{eq35} and \eqref{eq15}) $V^2$ as an
explicit function of ${\alpha}_i$  and using
def\/inition~\eqref{eq24}, we get f\/inally
\begin{gather}\label{eq36}
{\tanh}^2\vert{\alpha}_i\vert =\frac{{\tanh}^2{\alpha}_1(1 -
{\tanh}^2{\alpha}_2) + {\tanh}^2{\alpha}_2(1 -
{\tanh}^2{\alpha}_3) + {\tanh}^2{\alpha}_3(1 -
{\tanh}^2{\alpha}_1)}{1-{\tanh}^2{\alpha}_1{\tanh}^2{\alpha}_2{\tanh}^2{\alpha}_3}.
\end{gather}
This formula expresses the rapidity magnitude
$\vert{\alpha}_i\vert$ via rapidity components ${\alpha}_i$. As
should be in this case, $\vert{\alpha}_i\vert$ is a convex
function of its own arguments ${\alpha}_i$, which is invariant
with respect to their ref\/lection ${\alpha}_i\leftrightarrow
-{\alpha}_i.$ According to \eqref{eq36},
\begin{gather}\label{eq37}
\vert{\alpha}_i\vert\approx\sqrt{{{\alpha}_1}^2+{{\alpha}_2}^2+{{\alpha}_3}^2}
\end{gather}
at small $\vert{\alpha}_i\vert$ values. As $\vert{\alpha}_i\vert$
increases, the spheres \eqref{eq37} get deformed and transformed
at $\vert{\alpha}_i\vert\to\infty$ into the regular rhombic
dodecahedron shown in Fig.~\ref{fig1}.

To illustrate the above described transformation of spheres into a
regular rhombic dodecahedron, we shall consider f\/irst the
behavior of the section of surface $\vert{\alpha}_i\vert ={\rm
const}$  by plane ${\alpha}_3=0.$

The equation that describes the  section is
\begin{gather}\nonumber
\frac{{\alpha}_1+{\alpha}_2}{\sqrt{2}}= \pm \frac{1}{\sqrt{2}}\ln\bigg \{2\cosh{\vert{\alpha}_i\vert}-\cosh{\sqrt{2}\left (\frac{{\alpha}_1-{\alpha}_2}{\sqrt{2}}\right )} \\
\label{eq38} \phantom{\frac{{\alpha}_1+{\alpha}_2}{\sqrt{2}}=}{} +
\biggl [\left (2\cosh{\vert{\alpha}_i\vert}-\cosh{\sqrt{2}\left
(\frac{{\alpha}_1-{\alpha}_2}{\sqrt{2}}\right )}\right )^2-1\biggr
]^{1/2}\biggr \},
\end{gather}
with the variation limits for the argument of function
\eqref{eq38} being
\begin{gather*}
{\frac{{\alpha}_1-{\alpha}_2}{\sqrt{2}}}{\biggl\vert}_{{\alpha}_1+{\alpha}_2=0}=\mp\frac{1}{\sqrt{2}}\ln\left
\{2\cosh{\vert{\alpha}_i\vert}-1+\left [\left
(2\cosh{\vert{\alpha}_i\vert}-1\right )^2-1\right ]^{1/2}\right
\}.
\end{gather*}

\begin{figure}[t]

\centerline{\includegraphics[width=6.4cm]{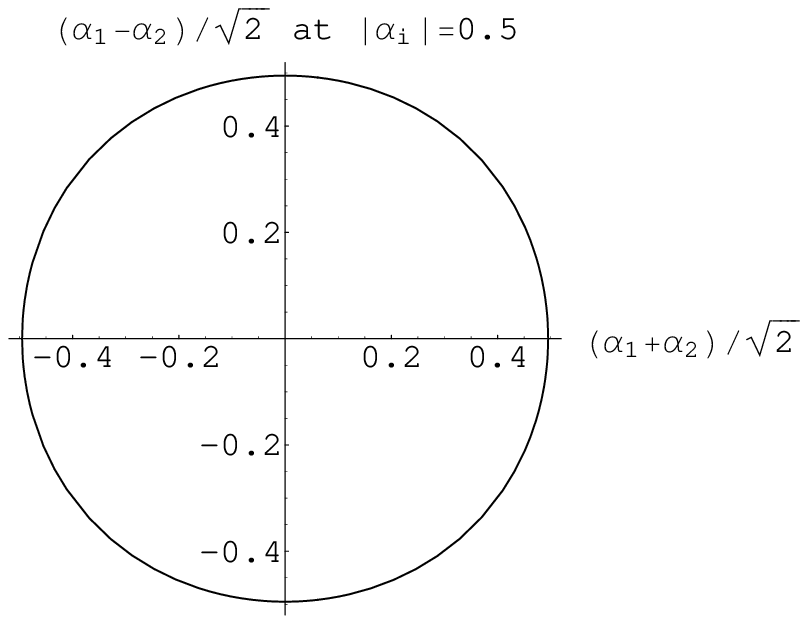}\qquad\quad
\includegraphics[width=6.4cm]{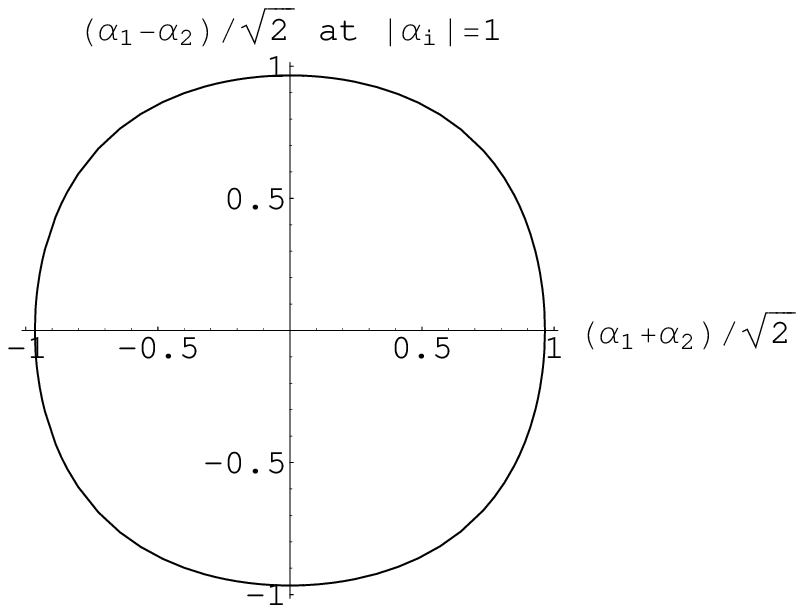}}

\bigskip

\centerline{\includegraphics[width=6.4cm]{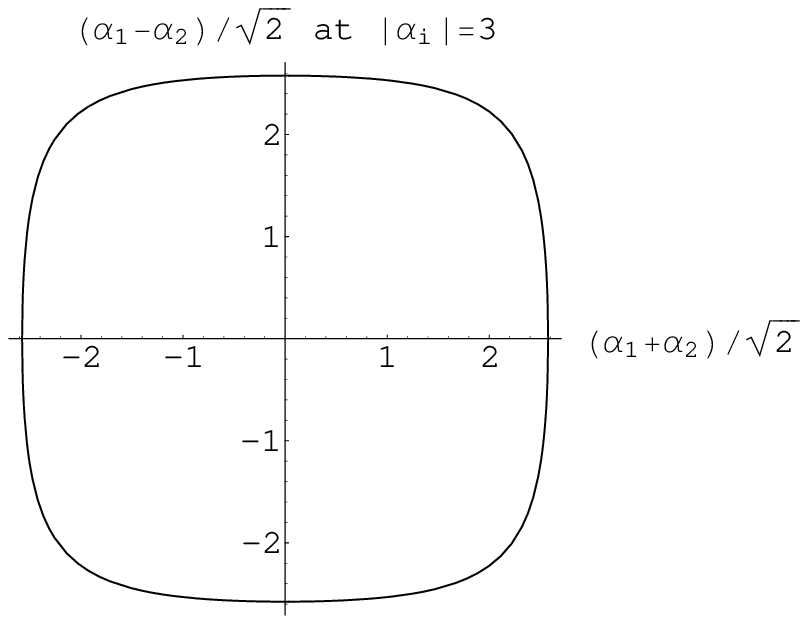}\qquad\quad
\includegraphics[width=6.4cm]{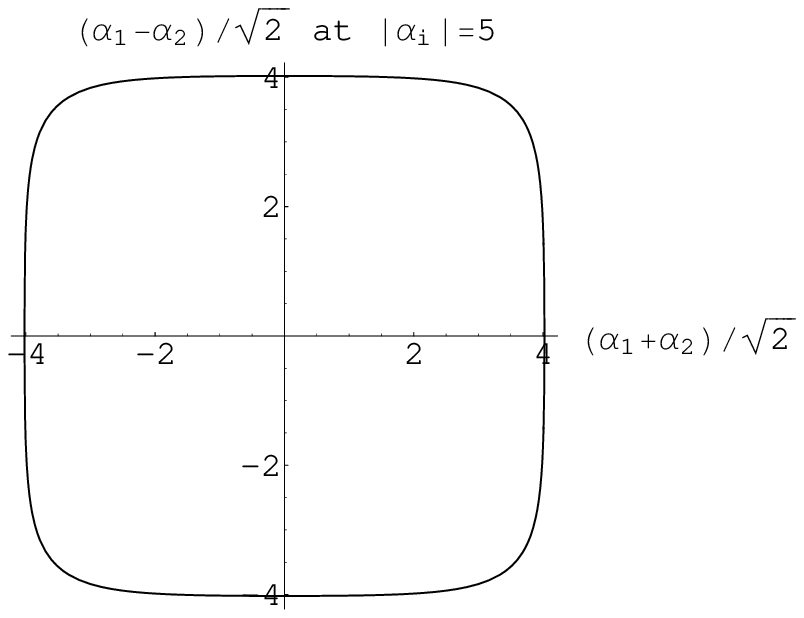}}

\vspace{-3mm}

  \caption{The sections of surfaces $\vert{\alpha _i}\vert =0.5$, $\vert{\alpha _i}\vert =1$, $\vert{\alpha _i}\vert =3 $ and $\vert{\alpha _i}\vert =5 $ by plane $\alpha _3=0,$ demonstrating the fact that, at $\vert{\alpha _i}\vert\to\infty$, the section of surface $\vert{\alpha _i}\vert ={\rm const}$ by the same plane tends to the section of the light front surface shown in Fig.~\ref{fig3}.}\label{fig2}
\vspace{-2mm}
 \end{figure}

\begin{figure}[t]
\centerline{\includegraphics[width=7.3cm]{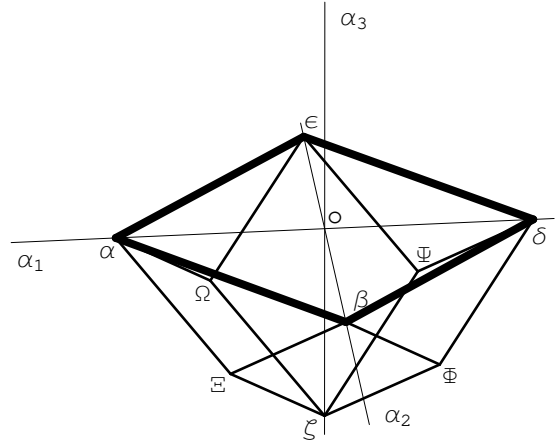}}
\vspace{-2mm}

\caption{Bold-faced square $\beta \alpha \epsilon \delta $ as
section of the light front surface (of the regular rhombic
dodecahedron) by plane $\alpha _3=0$.}\label{fig3}
\end{figure}

Incidentally, it should be noted that, when deriving \eqref{eq38}
we not only put ${\alpha}_3=0$ in the initial equation
\eqref{eq36}, but also introduced new variables
$({\alpha}_1+{\alpha}_2)/\sqrt{2}$ and
$({\alpha}_1-{\alpha}_2)/\sqrt{2}$ instead of ${\alpha}_1$ and
${\alpha}_2.$ The function \eqref{eq38} is plotted in
Fig.~\ref{fig2} at $\vert{\alpha}_i\vert =0.5$,
$\vert{\alpha}_i\vert =1$, $\vert{\alpha}_i\vert =3$ and
$\vert{\alpha}_i\vert =5.$ From the plots it is seen how, as
$\vert{\alpha}_i\vert$   increases, the circle gets transformed
gradually into the square $\beta \alpha \epsilon \delta $ shown in
Fig.~\ref{fig3}.

\begin{figure}[t]
\centerline{\includegraphics[width=6.3cm]{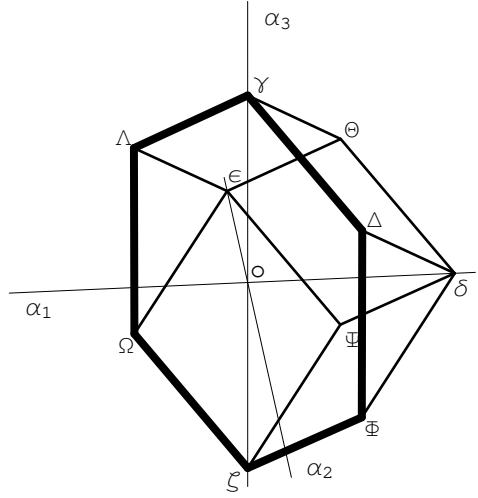}}
\vspace{-2mm}

\caption{Bold-faced hexagon $\Lambda \gamma \Delta \Phi \zeta
\Omega $ as section of the light front surface (of the regular
rhombic do\-deca\-hedron) by plane $(\alpha _1+\alpha
_2)=0$.}\label{fig4}
\end{figure}

Finally, to complete the picture, examine the behavior of the
section of surface $\vert{\alpha}_i\vert ={\rm const}$ by plane
$({\alpha}_1+{\alpha}_2)=0.$ The equation that describes the
section is
\begin{gather}\nonumber
\frac{{\alpha}_1-{\alpha}_2}{\sqrt{2}}= \pm \frac{1}{\sqrt{2}}\ln\biggl \{\bigl ({\cosh}^2{2{\alpha}_3}+2\cosh{2\vert{\alpha}_i\vert}+1\bigr )^{1/2}-\cosh{2{\alpha}_3} \\
\label{eq40} \phantom{\frac{{\alpha}_1-{\alpha}_2}{\sqrt{2}}=}{}
+\Bigl [\Bigl (\bigl
({\cosh}^2{2{\alpha}_3}+2\cosh{2\vert{\alpha}_i\vert}+1\bigr
)^{1/2}-\cosh{2{\alpha}_3}\Bigr )^2-1\Bigr ]^{1/2}\biggr \}.
\end{gather}
The variation limits for the argument of function \eqref{eq40} are
${{\alpha}_3}{\bigl\vert}_{{\alpha}_1-{\alpha}_2=0}=\mp\vert{\alpha}_i\vert$.

\begin{figure}[t]

\centerline{\includegraphics[width=6.4cm]{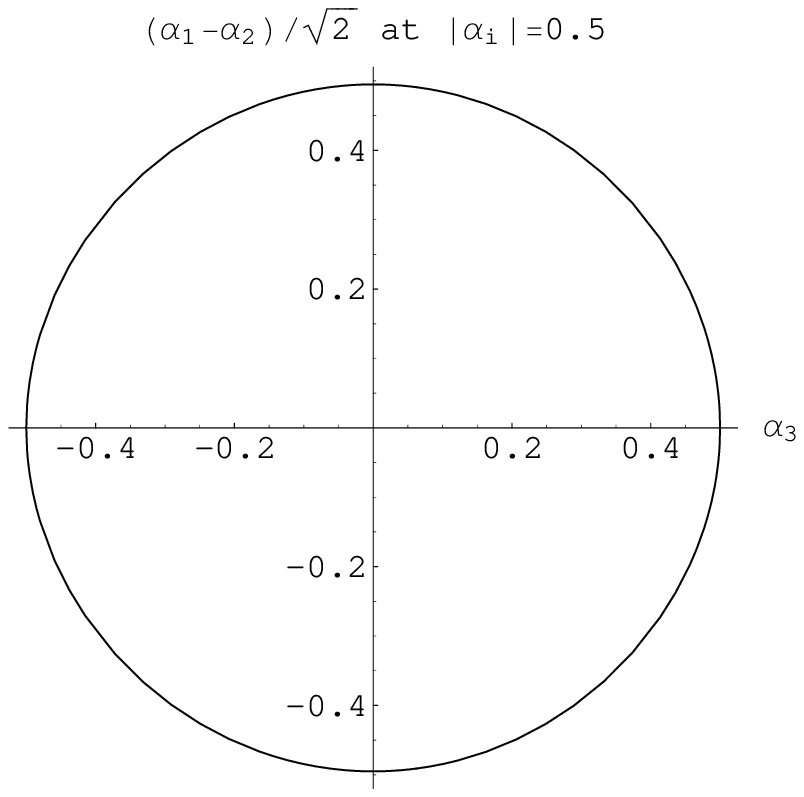}\qquad\quad
\includegraphics[width=6.4cm]{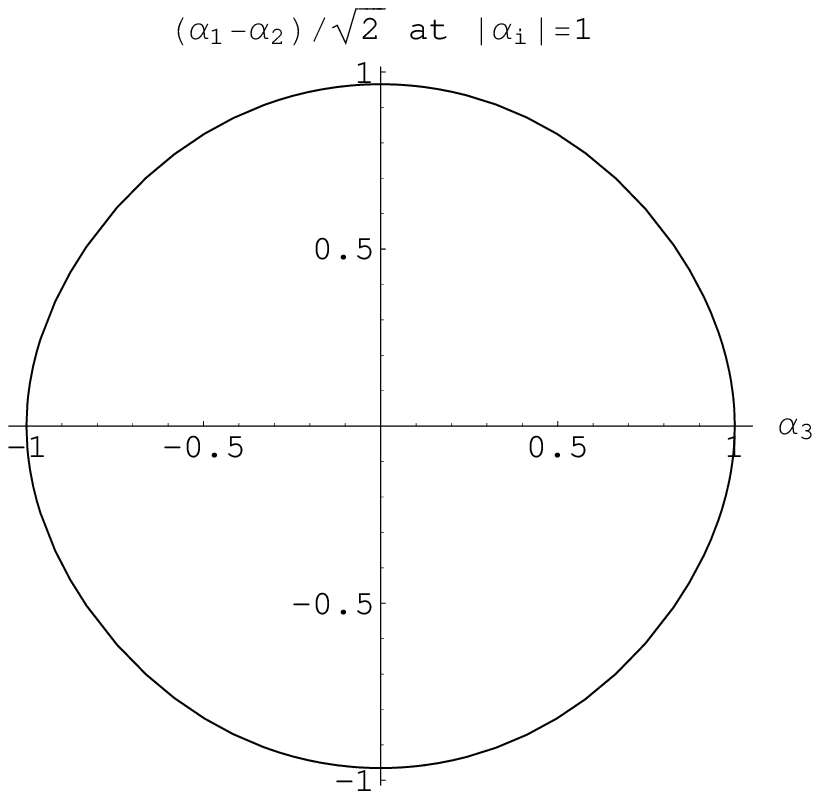}}

\bigskip

\centerline{\includegraphics[width=6.4cm]{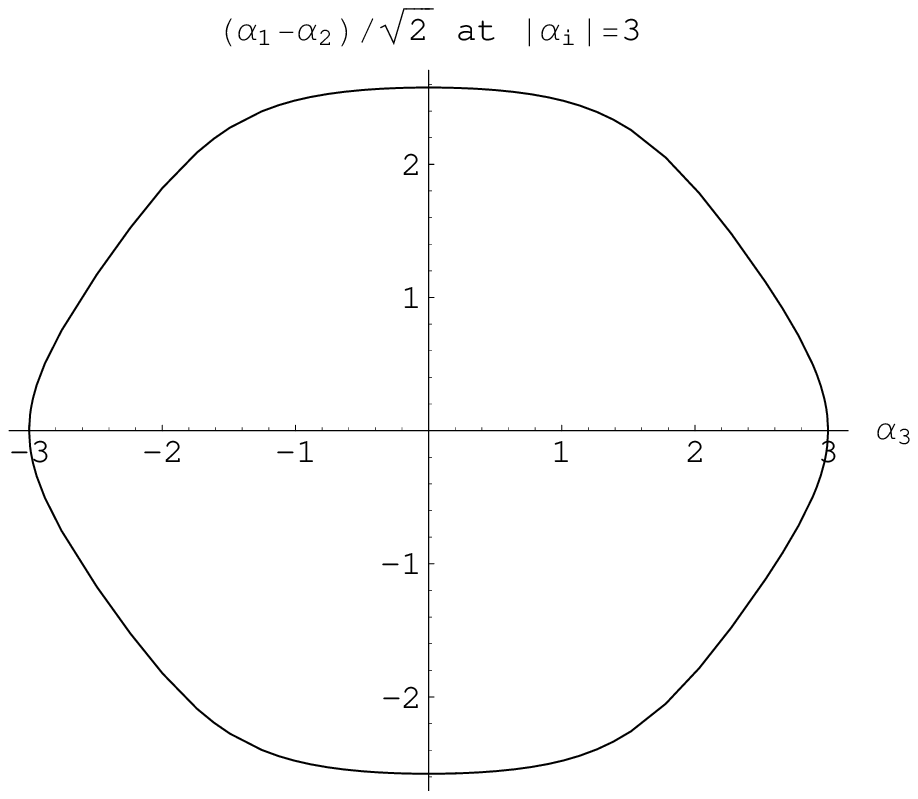}\qquad\quad
\includegraphics[width=6.4cm]{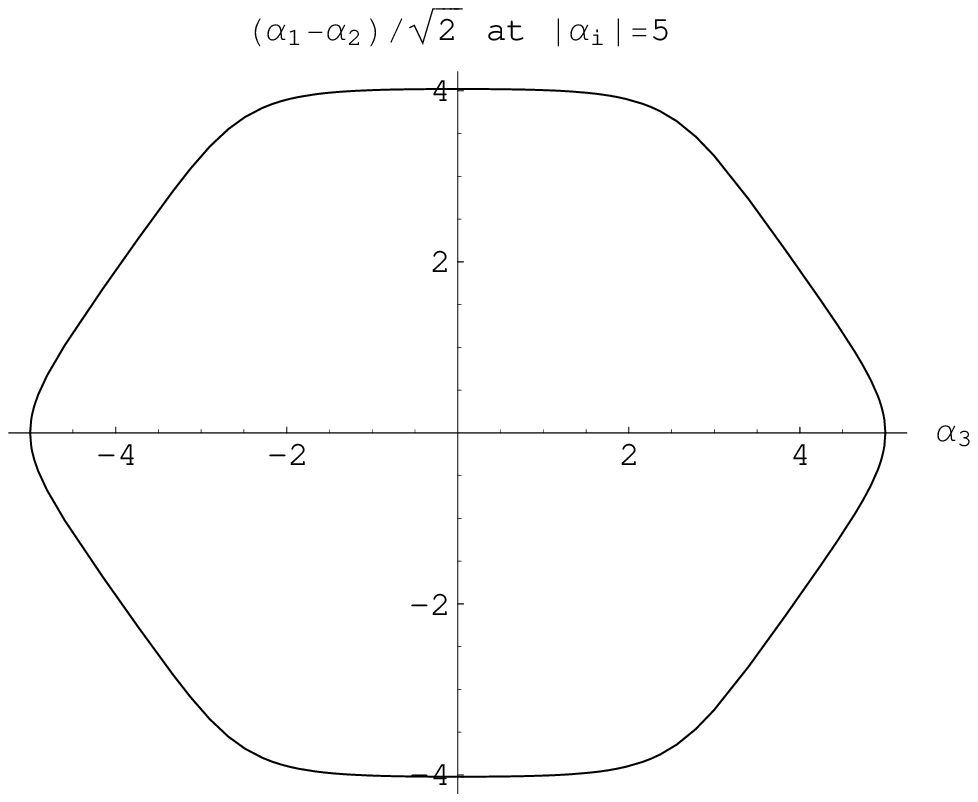}}

 \vspace{-3mm}

  \caption{The sections of surfaces $\vert{\alpha _i}\vert =0.5$, $\vert{\alpha _i}\vert =1$, $\vert{\alpha _i}\vert =3 $ and $\vert{\alpha _i}\vert =5 $ by plane $(\alpha _1+\alpha _2)=0$ demonstrating the fact that, at $\vert{\alpha _i}\vert\to\infty$, the section of surface $\vert{\alpha _i}\vert ={\rm const}$ by the same plane tends to the section of the light front surface shown in Fig.~\ref{fig4}.}\label{fig5}
 \end{figure}

It should be noted that, when deriving equation \eqref{eq40}, we
f\/irst introduced in \eqref{eq36}
$({\alpha}_1+{\alpha}_2)/\sqrt{2}$ and
$({\alpha}_1-{\alpha}_2)/\sqrt{2}$ instead of variables
${\alpha}_1$ and ${\alpha}_2$ and, after that, put
$({\alpha}_1+{\alpha}_2)=0.$ Fig.~\ref{fig5} shows the plots of
\eqref{eq40} at $\vert{\alpha}_i\vert =0.5$, $\vert{\alpha}_i\vert
=1$, $\vert{\alpha}_i\vert =3$ and $\vert{\alpha}_i\vert =5.$ From
the plots it is seen that, as~$\vert{\alpha}_i\vert$ increases,
the circle becomes gradually the hexagon $\Lambda \gamma \Delta
\Phi \zeta \Omega $ shown in Fig.~\ref{fig4}.

\section[A relation between the components of observable 3-velocity and its magnitude in the flat Finslerian event space with  entirely broken 3D isotropy]{A relation between the components of observable 3-velocity\\ and its magnitude in the f\/lat Finslerian event space\\ with  entirely broken 3D isotropy}

The Minkowski space domain of causally-related events that
correspond to motion of a classical particle is well-known to be
bounded by light cone surface. In this case, since the light cone
equation in orthogonal Galilean coordinates has the simplest
(canonical) form, use of the Einstein procedure of exchange of
light signals  leads to the trivial result, according to which the
magnitude of observable 3-velocity is expressed via the components
of the latter as $V^2=v_1^2+v_2^2+v_3^2.$ Besides, the components
$v_i$ of observable 3-velocity coincide in this case with the
respective components of coordinate 3-velocity, so that the range
of the physically permissible values of coordinate 3-velocity of a
classical particle is limited by the condition $0\leq V^2\leq 1,$
i.e.\ by a sphere of unit radius. It should be noted that the same
situation arises also in the case of Finslerian space with
partially broken 3D isotropy \eqref{eq1}. However, the range of
the physically permissible values of coordinate 3-velocity $v_i$
gets changed signif\/icantly when going to entirely anisotropic
Finslerian space~\eqref{eq10}.

\begin{figure}[t]

\centerline{\includegraphics[width=6.3cm]{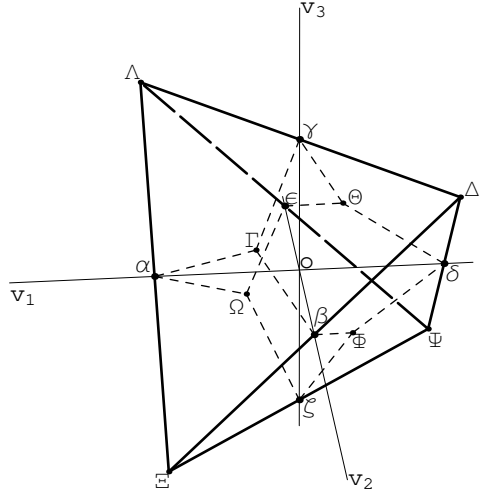}}
\vspace{-2mm}

\caption{The regular tetrahedron surface in the space of
coordinate velocities
             $v_i$ as an image of the light front surface.}\label{fig6}

\end{figure}

To obtain restrictions on the permissible $v_i$ values, Hubert
Goenner and me \cite{19bogoslovsky&goenner199899} made the valid
assumption that motion of a particle is in correspondence with
such events in space \eqref{eq10}, for which not only the
condition $dx_0>0,$ but also the set of inequalities
\begin{gather}
dx_0-dx_1-dx_2-dx_3\ge 0,\qquad
 dx_0-dx_1+dx_2+dx_3\ge 0,\nonumber\\
 dx_0+dx_1-dx_2+dx_3\ge 0,\qquad
 dx_0+dx_1+dx_2-dx_3\ge 0\label{eq42}
\end{gather}
are satisf\/ied, whence the following restrictions on   $v_i$ were
obtained
\begin{gather}\label{eq43}
1-v_1-v_2-v_3\ge 0,\\\label{eq44} 1-v_1+v_2+v_3\ge
0,\\\label{eq45} 1+v_1-v_2+v_3\ge 0,\\\label{eq46}
1+v_1+v_2-v_3\ge 0.
\end{gather}
According to \eqref{eq43}--\eqref{eq46}, the range of permissible
$v_i$ values is restricted by the surface of the regular
tetrahedron reproduced in Fig.~\ref{fig6}. It will be proved below
that, in this case the squared magnitude of the respective
observable 3-velocity~\eqref{eq35} fails to exceed unity, in which
case $V^2=1$ on the tetrahedron surface only. It is this fact that
allows us to  state that the domain bounded by the tetrahedron
surface serves as the domain of physically permissible coordinate
3-velocities of a classical particle. It should be noted also that
the conditions  \eqref{eq43}--\eqref{eq46}, which determine that
domain, are (at $dx_0>0$) the necessary and suf\/f\/icient
conditions for relations~\eqref{eq42} to be satisf\/ied. We can
conclude, therefore, that, from physical viewpoint, the
relations~\eqref{eq42} single out the domain of causally-related
events of event space~\eqref{eq10}, wherein the world line of any
classical particle resides. As to the boundary of that domain, it
is determined by the equations that have entered the set
\eqref{eq42}. Naturally, the boundary can  be called  the light
cone by analogy with the Minkowski space.

It can be readily verif\/ied that, when crossing the light cone,
we f\/ind ourselves in the domain of causally-unrelated events.
For that purpose, as clear from the above analysis, it is
suf\/f\/icient to verify that, when crossing the regular
tetrahedron surface in the space of coordinate 3-velocities~$v_i,$
we enter the domain of supraluminal velocities, i.e.\ we obtain
the values of squared observable 3-velocity~$V^2(v_i)$ that exceed
unity.

The function $V^2(v_i)$ behavior of interest for us can be found
out by examining its behavior on the set of all straight lines
that pass through the origin $o$ of the Cartesian coordinates
$v_i$ and, hence, intersect the surface of the regular
tetrahedron.

Considered as a typical example will be the behavior of $V^2(v_i)$
on a straight line that intersects the tetrahedron surface at
points $\Psi$ and $\Gamma$ (see Fig.~\ref{fig6}). According to
equations~\eqref{eq44}--\eqref{eq46}, point~$\Psi$ (tetrahedron
vertex) belongs to faces $\Psi\Xi\Lambda $, $\Psi\Delta\Xi$ and
$\Psi\Lambda \Delta $ and its coordinates are $v_1 = -1$,
$v_2=-1$, $v_3=-1,$ whereas equation~\eqref{eq43} determines the
coordinates $v_1=1/3$, $v_2=1/3$, $v_3=1/3$ of point $\Gamma,$
which is located at the center of the face $\Delta\Xi\Lambda $.
Therefore, the equations that describe the selected straight line
are
\begin{gather}\label{eq47}
v_1=t, \qquad v_2=t, \qquad v_3=t,
\end{gather}
where $t$ is the parameter that characterizes the location of a
point on the given line. In particular, $t=-1$ corresponds to
point $\Psi,$ $t=0$ to point $o$, and $t=1/3$ to point $\Gamma.$
Substituting~\eqref{eq47} in~\eqref{eq35}, we obtain $V^2$ in the
form of the following explicit function of running  parameter $t$
\begin{gather*}
V^2(t)=3t^2/(1-2t).
\end{gather*}
It is quite obvious now that
\begin{enumerate}\itemsep=0pt
\item[i)] at the center $o$ of the tetrahedron, i.e.\ at $t=0,$
$V^2(0)=0$; \item[ii)] as $t$ increases, i.e. as we approach the
tetrahedron surface, $V^2(t)$ increases steadily, so that at
$t=1/3$ (i.e. at point $\Gamma$ on the tetrahedron surface)
$V^2(t)$ takes on value $V^2(1/3)=1$; \item[iii)] as $t$ increases
further, i.e. as we move away from  the tetrahedron, $V^2(t)$
keeps increasing steadily and
$\lim\limits_{t\to{1/2}}V^2(t)=\infty$; \item[iv)] at $t>1/2,$ we
appear in the region where the squared observable 3-velocity takes
on negative values, i.e.\ $V^2(t)<0$; \item[v)] as $t$ decreases
(from $t=0$), i.e. as we move away from point $o$ (where $V^2=0$)
and, hence, as we approach the tetrahedron surface again, $V^2(t)$
increases similarly to the case ii) and, at $t=-1$, i.e.\ at point
$\Psi$ on the tetrahedron surface,  takes on value
$V^2(-1)=V^2(1/3)=1$; \item[vi)] as $t$ decreases further, i.e. as
we move away from  the tetrahedron again, $V^2(t)$ keeps
increasing steadily, with $V^2(t)\approx{-3t/2}$ at $t\ll{-1}$.
\end{enumerate}

The method proposed  above can be used to study the behavior of
function $V^2(v_i)$ throughout the space of coordinate velocities
$v_i$ and (which is of particular importance for quantum theory),
to accurately determine the domains in the space of $v_i$ where
$V^2>1$ or $V^2<0$. The two domains correspond to propagation of
virtual particles only, the f\/irst domain corresponding to
causally-unrelated (tachyonic) events in space~\eqref{eq10}.
Within the scope of this work, however, we shall limit ourselves
to the above analysis. The point is that the behavior of function
$V^2(v_i)$ along straight line \eqref{eq47} is a good illustration
of both the specif\/ic features of its behavior throughout the
external (with respect to the tetrahedron) domain and the fact
that $0\le V^2\le 1$ in the domain belonging to the tetrahedron,
i.e. just in the domain of $v_i,$  which is of interest to us in
this work.

Return now to Fig.~\ref{fig6}. Since the regular tetrahedron
surface is an image of the light front surface in the space of
$v_i,$ all the initial light rays and their respective ref\/lected
rays divide the tetrahedron into six pairs of mutually conjugate
sectors
\begin{gather*}
\Gamma \gamma \Delta \beta o\ \longleftrightarrow \ \Psi \zeta
\Omega \epsilon o,\qquad \Gamma \beta \Xi \alpha o\
\longleftrightarrow \ \Psi \epsilon
\Theta \delta o,\\
\Gamma \alpha \Lambda \gamma o \ \longleftrightarrow \  \Psi
\delta \Phi \zeta o,\qquad \Omega \epsilon \Lambda \alpha o \
\longleftrightarrow \  \Delta
\beta \Phi \delta o,\\
\Omega \alpha \Xi \zeta o \ \longleftrightarrow \  \Delta \delta
\Theta \gamma o, \qquad \Theta \gamma \Lambda \epsilon o \
\longleftrightarrow \  \Xi \zeta \Phi \beta o.
\end{gather*}
Therefore, the Einstein procedure of exchange of light signals
makes it possible to obtain only the respective sectorial
components of observable 3-velocity (see
\cite{19bogoslovsky&goenner199899})
\begin{gather*}
v_{\Gamma \gamma\Delta \beta o}=(v_3+v_2)/(1-v_1), \qquad
v_{\Psi \zeta \Omega \epsilon o}=-(v_3+v_2)/(1-v_1),\\
v_{\Gamma \beta \Xi \alpha o}=(v_1+v_2)/(1-v_3), \qquad
v_{\Psi \epsilon \Theta \delta o}=-(v_1+v_2)/(1-v_3),\\
v_{\Gamma \alpha \Lambda \gamma o}=(v_1+v_3)/(1-v_2), \qquad
v_{\Psi \delta \Phi \zeta o}=-(v_1+v_3)/(1-v_2),\\
v_{\Omega \epsilon \Lambda \alpha o}=(v_1-v_2)/(1+v_3), \qquad
v_{\Delta \beta \Phi \delta o}=-(v_1-v_2)/(1+v_3),\\
v_{\Omega \alpha \Xi \zeta o}=(v_1-v_3)/(1+v_2), \qquad
v_{\Delta \delta \Theta \gamma o}=-(v_1-v_3)/(1+v_2),\\
v_{\Theta \gamma \Lambda \epsilon o}=(v_3-v_2)/(1+v_1), \qquad
v_{\Xi \zeta \Phi \beta o}=-(v_3-v_2)/(1+v_1),
\end{gather*}
where $v_i$ are the coordinate 3-velocity components. As to the
relation between the magnitude of observable 3-velocity and its
sectorial components, we shall derive it as follows.

First, using \eqref{eq15} and simplifying the notations in quite
an evident way, we represent twelve (actually six) sectorial
components of observable 3-velocity as explicit functions of
${\alpha}_i$
\begin{gather}
v_{\Gamma \gamma\Delta \beta o}\equiv v_{3+2}=\tanh (\alpha
_3+\alpha _2),\qquad
v_{\Gamma \beta \Xi \alpha o}\equiv v_{1+2}=\tanh (\alpha _1+\alpha _2),\nonumber\\
v_{\Gamma \alpha \Lambda \gamma o}\equiv v_{1+3}=\tanh (\alpha
_1+\alpha _3),\qquad
v_{\Omega \epsilon \Lambda \alpha o}\equiv v_{1-2}=\tanh (\alpha _1-\alpha _2),\nonumber\\
v_{\Omega \alpha \Xi \zeta o}\equiv v_{1-3}=\tanh (\alpha
_1-\alpha _3),\qquad v_{\Theta \gamma \Lambda \epsilon o}\equiv
v_{3-2}=\tanh (\alpha _3-\alpha _2).\label{eq49}
\end{gather}
Besides, it is expedient to represent formula \eqref{eq36} as
\begin{gather}
{\tanh}^2\vert{\alpha}_i\vert =1 - 8\big [2+\cosh{2(\alpha_1+\alpha_2)}+\cosh{2(\alpha_1-\alpha_2)}+\cosh{2(\alpha_3+\alpha_2)}\nonumber\\
\phantom{{\tanh}^2\vert{\alpha}_i\vert =}{}
+\cosh{2(\alpha_3-\alpha_2)}+\cosh{2(\alpha_1+\alpha_3)}+\cosh{2(\alpha_1-\alpha_3)}\big]^{-1}.\label{eq50}
\end{gather}
Allowing for \eqref{eq24} and using \eqref{eq50} and \eqref{eq49},
we obtain
\begin{gather}
V^2=1 -8 \left [2 + \frac{1 + v^{2}_{1+2}}{1 - v^{2}_{1+2}} +
\frac{1 + v^{2}_{1+3}}{1 - v^{2}_{1+3}} + \frac{1 + v^{2}_{3+2}}{1
- v^{2}_{3+2}} +
\frac{1 + v^{2}_{1-2}}{1 - v^{2}_{1-2}} \right.\nonumber\\
\left. \phantom{V^2=}{} + \frac{1 + v^{2}_{1-3}}{1 - v^{2}_{1-3}}
+ \frac{1 + v^{2}_{3-2}}{1 - v^{2}_{3-2}}\right
]^{-1}.\label{eq51}
\end{gather}

According to \eqref{eq51}, the square of observable 3-velocity is
a symmetric function of its six observable sectorial components
\eqref{eq49}. It should be noted that these components  are
constrained by three relations and, hence, are not independent. It
can be readily verif\/ied that, for example,
\begin{gather}\label{eq52}
v_{1-2}=\frac{v_{1+3}-v_{3+2}}{1-v_{1+3}v_{3+2}},\qquad
v_{1-3}=\frac{v_{1+2}-v_{3+2}}{1-v_{1+2}v_{3+2}},\qquad
v_{3-2}=\frac{v_{1+3}-v_{1+2}}{1-v_{1+3}v_{1+2}}.
\end{gather}
Therefore, any three dif\/ferent observable sectorial components
taken from the set \eqref{eq49} serve as one of the possible
non-orthogonal basises. For example, having chosen the sectorial
compo\-nents~$v_{1+2}$, $v_{1+3}$, $v_{3+2}$ to be a particular
basis, we can use \eqref{eq51} and \eqref{eq52} to obtain
\begin{gather}
V^2=1-\big\{(1+v_{1+2}v_{1+3}+v_{1+2}v_{3+2}+v_{1+3}v_{3+2})^2\nonumber\\
\phantom{V^2=}{} -
(v_{1+2}+v_{1+3}+v_{3+2}+v_{1+2}v_{1+3}v_{3+2})^2\big\}
\nonumber\\
\phantom{V^2=}{}
\times\big\{1-(v_{1+2}+v_{1+3}+v_{3+2}-v_{1+2}v_{1+3}v_{3+2})^2/4\big\}^{
-1}.\label{eq53}
\end{gather}
If, in particular, $v_{1+3}\not= 0$, while $v_{1+2}=v_{3+2}=0,$
then
\begin{gather}\label{eq54}
V^2=\frac{3v^{2}_{1+3}}{4-v^{2}_{1+3}}.
\end{gather}
The fact that the square of observable 3-velocity does not
coincide (in virtue of \eqref{eq54}) with the square of its sole
(non-zero) component $v_{1+3}$    indicates that the given basis
is non-orthogonal. On going to, for example, another
non-orthogonal basis $v_{1-2}$, $v_{1+3}$, $v_{3-2}$ (via
\eqref{eq52}) we get $v_{1-2}=v_{1+3}, v_{1+3}=v_{1+3},
v_{3-2}=v_{1+3}.$ In this case $v_{1+2}=v_{1-3}=v_{3+2}=0,$ so
\eqref{eq51} leads to $V^2$ expressed again via \eqref{eq54}. It
should be noted that, since $v_{1+3}$  is an observable sectorial
velocity, the maximum value of its square is $v^{2}_{1+3}=1.$ In
this case, as expected, formula \eqref{eq54} gives $V^2=1.$

The situation with  observable 3-velocities in Finslerian event
space \eqref{eq10} can be clarif\/ied f\/inally by examining the
3-velocities in non-relativistic limit, i.e.
\begin{gather}\label{eq55}
\vert{v_{1+2}}\vert\ll 1, \qquad \vert{v_{1+3}}\vert\ll 1, \qquad
\vert{v_{3+2}}\vert\ll 1.
\end{gather}
In this case \eqref{eq53} turns into
\begin{gather}\label{eq56}
V^2=\left [3\left (v^{2}_{1+2}+v^{2}_{1+3}+v^{2}_{3+2}\right
)/2-v_{1+2}v_{1+3}-v_{1+2}v_{3+2}-v_{1+3}v_{3+2}\right ] /2.
\end{gather}
By its meaning, \eqref{eq56} is a positively determined quadratic
form, so it can be reduced to a sum of squares. The explicit form
of the appropriate transformation can be found readily. To that
end, the f\/irst three relations of \eqref{eq49} and relations
\eqref{eq15} are suf\/f\/icient to use with due allowance for
\eqref{eq55}. As a result we get
\begin{gather}\label{eq57}
v_{1+2}=v_1+v_2, \qquad v_{1+3}=v_1+v_3, \qquad v_{3+2}=v_3+v_2.
\end{gather}
Accordingly, the inverse relations that represent the components
$v_i$ of coordinate 3-velocity appear as
\begin{gather*}
v_1 = (v_{1+3} + v_{1+2} - v_{3+2})/2, \qquad v_2 = (v_{1+2} +
v_{3+2} - v_{1+3})/2, \\ v_3 = (v_{1+3} + v_{3+2} - v_{1+2})/2.
\end{gather*}
On substituting \eqref{eq57} in \eqref{eq56}, we see that, in the
non-relativistic limit, the squared observable 3-velocity equals a
sum of squared components of the coordinate 3-velocity, i.e.\ $V^2
= {v_1}^2 +{v_2}^2 +{v_3}^2$.

\section{Conclusion}

As early as our works \cite{19bogoslovsky&goenner199899}, where
the  entirely anisotropic Finslerian event space  was f\/irst
treated, Hubert Goenner and the author  encountered the unusual
situation in which the Einstein procedure of exchange of light
signals failed to properly determine the magnitude of the
observable 3-velocity and was only able to determine the
components of the observable 3-velocity. As a~result, the
necessity arose that an algorithm  be constructed to permit
determination of the  magnitude of the observable 3-velocity,
thereby formalizing the condition that singles out the domain of
causally-related events in space \eqref{eq10}. The said events are
characterized by the fact that the respective magnitudes of
observable 3-velocities do not exceed the speed of light.

Although constructing the above-mentioned algorithm as realized in
the present work was partly heuristic (the identif\/ication of
functions \eqref{eq33} and \eqref{eq34} to be those exhibiting
special transformational properties is meant), the comprehensive
analysis of the f\/inal formula \eqref{eq35} has shown that it
determines  the magnitude of observable 3-velocity correctly.
Besides, according to~\eqref{eq51}, formula \eqref{eq35} has led
to quite a reasonable relation between the components of
observable 3-velocity and its magnitude. Thus, in addition to the
isometry group \eqref{eq12}, we have now at our disposal another
important tool to study the behavior of fundamental f\/ields in
entirely anisotropic Finslerian space \eqref{eq10}. The respective
studies, as well as the explicit form of the fundamental f\/ield
equations generalized to meet the requirement of invariance with
respect to group \eqref{eq12}, will be published elsewhere. As to
the present work, to avoid any possible misunderstandings, the
following is expedient to note.

It is common knowledge  that, in the conventional relativistic
quantum f\/ield theory, 4-momen\-tum ($p_0,\boldsymbol p$)
conjugated canonically to orthogonal Galilean coordinates
($x_0,\boldsymbol x$) is more fundamental than 3-velocity
$\boldsymbol v.$ At the same time, in the case of a free
single-particle state with certain 4-momentum $|p_0,\boldsymbol
p\rangle,$ the 3-velocity as a quantum observable has certain
value $\boldsymbol v=\boldsymbol p/p_0$ that coincides with
classical observable 3-velocity. In the presence of Lorentz
violation, however, the connection between the momentum and the
velocity can be highly nontrivial, the fact that was f\/irst
demonstrated by Don Colladay and Alan Kosteleck\'y in the original
SME paper in \cite{2colladay&kostelecky}. Attention should be paid
to the fact that a similar ef\/fect occurs also within the
framework of relativistically invariant Finslerian approach to the
Lorentz symmetry violation. In the case of f\/lat axially
symmetric Finslerian event space \eqref{eq1}, for instance, for a
free single-particle state with certain 4-momentum
$|p_0,\boldsymbol p\rangle$, the 3-velocity as a quantum
observable has certain value
\begin{gather*}
{\boldsymbol v} = \frac{(1+r)(p_0 - {\boldsymbol p}{\boldsymbol
\nu}) {\boldsymbol p} - r (p_0^2 - {\boldsymbol p}^2) {\boldsymbol
\nu}}{(1+r)(p_0 - {\boldsymbol p}{\boldsymbol \nu})p_0 - r(p_0^2 -
{\boldsymbol p}^2)}
\end{gather*}
that coincides with classical observable 3-velocity in the said
event space (see the original  paper~\cite{4bogoslovsky1977}).

In the case of f\/lat entirely anisotropic Finslerian event space
\eqref{eq10} the above notion of 3-velocity involves subtleties
associated with the fact that 4-momentum ($p_0, p_i$) is
canonically conjugate to {\it{nonorthogonal}} Galilean coordinates
($x_0, x_i$) used to prescribe metric \eqref{eq10}. For a~free
single-particle state with certain 4-momentum $|p_0, p_i\rangle$,
as a result, the 3-velocity $v_i$ as a~quantum observable has the
certain value, which is the solution for the set of algebraic
equations
\begin{gather*}
a_{qi}v_{i}=b_{q},
\end{gather*}
where
\begin{gather*}
a_{11}= a_{12} =(p_0+p_3)(1+r_3)-(p_1+p_2)(r_1+r_2),\\
a_{13}= b_{1} =(p_1+p_2)(1+r_3)-(p_0+p_3)(r_2+r_3),\\
a_{21}= - b_{2} =(p_0-p_1)(r_2-r_3)-(p_2-p_3)(1-r_1),\\
a_{22}= - a_{23} =(p_0-p_1)(1-r_1)-(p_2-p_3)(r_2-r_3),\\
a_{31}= b_{3} =(p_2+p_3)(1+r_1)-(p_0+p_1)(r_2+r_3),\\
a_{32}= a_{33} =(p_0+p_1)(1+r_1)-(p_2+p_3)(r_2+r_3).
\end{gather*}
In the given case, thus, the  value of $v_i$ taken as a quantum
observable coincides with classical {\it{coordinate}} 3-velocity
(see \cite{32bogoslovsky2007}). This implies that
relations~\eqref{eq35} and~\eqref{eq49} should be also used to
determine the magnitude and the components of the
{\it{observable}} 3-velocity that would correspond to the free
single-particle state with certain 4-momentum $|p_0, p_i\rangle$.

It should be emphasized that the single-particle states with
certain 4-momentum occur in either of the types of f\/lat
anisotropic Finslerian spaces, with pure kinetic part of the
4-momentum being in principle impossible to separate from the
latter. This fact indicates that any interaction induced by one or
another anisotropy of the homogeneous medium that f\/ills
space~\eqref{eq1} or space~\eqref{eq10} is nonseparable, while the
above-mentioned single-particle states are the collective
excitations (quasi-particles) that exist in either medium.

The most likely candidate as entirely anisotropic medium that
creates  anisotropy of space~\eqref{eq10} is a three-gluon
condensate. A feasibility of forming the three-gluon condensate
ensues from work \cite{33arbuzov2007} that demonstrates a
spontaneous generation of three-gluon gauge invariant ef\/fective
interaction. Note in the interim that the inert mass tensor for
nonrelativistic quasi-particles in entirely anisotropic
crystal-like medium is determined by formula \eqref{eq11}.

As to  axially symmetric anisotropic medium that creates
anisotropy of space~\eqref{eq1}, an axially symmetric
fermion-antifermion condensate simulates this sort of
crystal-like medium. The condensate appears as a vacuum solution
of the Finslerian theory of fermions, whose Lagrangian~\eqref{eq9}
is constructed proceeding from the requirement of invariance with
respect to group DISIM${_b}$(2). After the appropriate shift and
expansion, the theory~\eqref{eq9} gets quite def\/ined, so the
condensate stability problem can be treated within the framework
of that theory. This problem is important to solve just because
the existing stringent experimental bounds on the space anisotropy
magnitude may be indicative of feasible evaporation of the axially
symmetric fermion-antifermion condensate, resulting in the
present-day very small space anisotropy. If the condensate
evapo\-ration assumption proves to be correct, then the question
posed in~\cite{16gibbons&gomis&pope2007}, namely, why is the
anisotropy so small, may be answered.

Strictly speaking, the condensate stability problem must be solved
in terms of such an extension of the theory~\eqref{eq9} that would
include couplings to other f\/ields, to gravity f\/ield in
particular. Although a possibility does exist in principle for the
theory~\eqref{eq9} to be extended appropriately within the
framework of Finslerian approach to the Lorentz symmetry
violation, realizing such a plan falls outside the scope of the
present work. It should be noted in conclusion, nevertheless, that
the f\/irst notable attempt of describing the anisotropy-gravity
interaction in terms of the Finslerian model for curved space-time
was made in \cite{8bogoslovsky199394}. We cannot but mention also
work~\cite{11kostelecky2004}, wherein Alan Kosteleck\'y
demonstrated that gravity with explicit Lorentz violation could
not be described by Riemann (or even Riemann--Cartan) geometry and
suggested on his own a~Finslerian origin for it instead.

\subsection*{Acknowledgements}

The author is indebted to Boris Arbuzov for informative discussion
concerning the feasibility of forming the three-gluon condensate.
The author is also thankful to Gary Gibbons for a helpful
discussion of the status of the present-day experimental upper
bounds on the space anisotropy magnitude and the prospects of the
relevant fresh experiments. Finally, the author expresses his
gratitude to Dimitri Pavlov and Grigori Garas'ko for the inspiring
discussions that stimulated preparing the present work.

Separately, the author would like to thank Hubert Goenner for
many-year fruitful collaboration and also the Referees
for their valuable remarks that have permitted the paper to be
much improved.

This work was supported in part by the Russian Foundation for
Basic Research under grant RFBR-07-01-91681-RA\_a and by the
Non-Commercial Foundation for Finsler Geometry Research.

\pdfbookmark[1]{References}{ref}
\LastPageEnding


\begin{thebibliography}{99}

\footnotesize\itemsep=0pt

\bibitem{1kostelecky&samuel} Kosteleck\'y~A., Samuel~S., Spontaneous breaking of Lorentz symmetry in string theory, {\it Phys.~Rev.~D} {\bf 39} (1989), 683--685.

\bibitem{2colladay&kostelecky} Colladay~D., Kosteleck\'y~A., CPT violation
and the standard model, {\it Phys.~Rev.~D} {\bf 55} (1997), 6760--6774, \href{http://arxiv.org/abs/hep-ph/9703464}{hep-ph/9703464}.\\
Colladay~D., Kosteleck\'y~A., Lorentz-violating extension of the
standard model, {\it Phys.~Rev.~D} {\bf 58} (1998), 116002,
23~pages,
\href{http://arxiv.org/abs/hep-ph/9809521}{hep-ph/9809521}.

\bibitem{3kosteleckyEd} Kosteleck\'y A. (Editor), CPT and Lorentz symmetry III,
Singapore, World Scientif\/ic, 2005.\\
 Kosteleck\'y A. (Editor), CPT and Lorentz symmetry IV,
Singapore, World Scientif\/ic, 2008.

\bibitem{4bogoslovsky1977} Bogoslovsky~G.Yu., A special relativistic theory
of the locally anisotropic space-time, {\it Nuovo Cimento~B} {\bf 40} (1977), 99--134,
Erratum,  {\it Nuovo Cimento~B} {\bf 43} (1978), 377--378.

\bibitem{5tavakol1985} Tavakol~R., van den Bergh~N., Finsler spaces and the underlying geometry of space-time, {\it Phys.~Lett.~A} {\bf 112} (1985),  23--25.

\bibitem{6tavakol1986} Tavakol~R., van den Bergh~N., Viability criteria for the theories of gravity and Finsler spaces, {\it Gen. Relativity Gravitation} {\bf 18} (1986), 849--859.

\bibitem{7bogoslovskyBook} Bogoslovsky~G.Yu., Theory of locally anisotropic space-time, Moscow, Moscow Univ.~Press, 1992.

\bibitem{8bogoslovsky199394} Bogoslovsky~G.Yu., Finsler model of space-time, {\it Phys. Part. Nucl.} {\bf 24} (1993), 354--379.\\
 Bogoslovsky~G.Yu.,     A viable model of locally anisotropic space-time and the Finslerian generalization of the relativity theory, {\it Fortschr. Phys.} {\bf 42} (1994), 143--193.

\bibitem{9bogoslovsky&goenner2004} Bogoslovsky~G.Yu., Goenner~H.F., Concerning the generalized Lorentz symmetry and the  generalization of the Dirac equation, {\it Phys.~Lett.~A} {\bf 323} (2004), 40--47, \href{http://arxiv.org/abs/hep-th/0402172}{hep-th/0402172}.

\bibitem{10bogoslovsky2005} Bogoslovsky~G.Yu., Subgroups of the group of generalized Lorentz transformations and their geometric invariants, {\it SIGMA} {\bf 1} (2005), 017, 9~pages, \href{http://arxiv.org/abs/math-ph/0511077}{math-ph/0511077}.\\
 Bogoslovsky~G.Yu.,   Lorentz symmetry violation without violation of relativistic symmetry, {\it Phys.~Lett.~A} {\bf 350} (2006), 5--10, \href{http://arxiv.org/abs/math-ph/0511077}{hep-th/0511151}.

\bibitem{11kostelecky2004} Kosteleck\'y~A., Gravity, Lorentz violation, and the Standard Model, {\it Phys.~Rev.~D} {\bf 69} (2004),  105009,  20~pages, \href{http://arxiv.org/abs/hep-th/0312310}{hep-th/0312310}.

\bibitem{12bailey&kostelecky2006} Bailey~Q.G., Kosteleck\'y~A., Signals for Lorentz violation in post-Newtonian gravity, {\it Phys.~Rev.~D} {\bf 74} (2006),  045001, 46~pages, \href{http://arxiv.org/abs/gr-qc/0603030}{gr-qc/0603030}.

\bibitem{13girelly&liberati&sindoni2007} Girelly~F., Liberati~S., Sindoni~L., Planck-scale modif\/ied dispersion relations and Finsler geometry, {\it Phys.~Rev.~D} {\bf 75} (2007), 064015, 9~pages, \href{http://arxiv.org/abs/gr-qc/0611024}{gr-qc/0611024}.

\bibitem{14ghosh&pal2007} Ghosh~S., Pal~P., Deformed special relativity and deformed
symmetries in a canonical framework, {\it Phys.~Rev.~D} {\bf 75}
(2007), 105021, 11~pages,
\href{http://arxiv.org/abs/hep-th/0702159}{hep-th/0702159}.

\bibitem{15bogoslovsky2007} Bogoslovsky~G.Yu., Some physical displays of the space anisotropy relevant to the feasibility of its being detected at a laboratory, \href{http://arxiv.org/abs/0706.2621}{arXiv:0706.2621}.

\bibitem{16gibbons&gomis&pope2007} Gibbons~G.W., Gomis Joaquim, Pope~C.N., General very special relativity is Finsler geometry, {\it Phys.~Rev.~D} {\bf 76} (2007), 081701(R), 5~pages, \href{http://arxiv.org/abs/0707.2174}{arXiv:0707.2174}.\\
Cohen~A.G., Glashow~S.L., Very special relativity, {\it
Phys.~Rev.~Lett.} {\bf 97} (2006), 021601, 3~pages,
\mbox{\href{http://arxiv.org/abs/hep-ph/0601236}{hep-ph/0601236}}.

\bibitem{17mavromatos2007} Mavromatos~N.E., Lorentz invariance violation from string theory, \href{http://arxiv.org/abs/0708.2250}{arXiv:0708.2250}.

\bibitem{18sindoni2007} Sindoni~L., The Higgs mechanism in Finsler spacetimes,
\href{http://arxiv.org/abs/0712.3518}{arXiv:0712.3518}.

\bibitem{19bogoslovsky&goenner199899} Bogoslovsky~G.Yu., Goenner~H.F., On the possibility of phase transitions in the geometric structure of space-time, {\it Phys.~Lett.~A} {\bf 244} (1998), 222--228, \href{http://arxiv.org/abs/gr-qc/9804082}{gr-qc/9804082}.\\
 Bogoslovsky~G.Yu., Goenner~H.F.,     Finslerian spaces possessing local relativistic symmetry,
{\it Gen. Relativity Gravitation} {\bf 31} (1999), 1565--1603,
\href{http://arxiv.org/abs/gr-qc/9904081}{gr-qc/9904081}.

\bibitem{20bogoslovsky1973} Bogoslovsky~G.Yu., On a special relativistic theory of anisotropic space-time, {\it Dokl.~Akad.~Nauk~SSSR} {\bf 213} (1973), 1055--1058.

\bibitem{21patera&winternitz&zassenhaus1975} Patera~J., Winternitz~P.,
Zassenhaus~H., Continuous subgroups of the fundamental groups of
physics. II. The similitude group, {\it J.~Math.~Phys.} {\bf 16}
(1975), 1615--1624.

\bibitem{22wintwrnitz&fris1965} Winternitz~P., Fri\v{s}~I., Invariant
expansions of relativistic amplitudes and subgroups of the proper
Lorentz group, {\it Yadern.~Fiz.} {\bf 1} (1965), 889--901.

\bibitem{23bogoslovsky1982} Bogoslovsky~G.Yu., The proper time, spatial distances and clock synchronization in the locally anisotropic space-time,
{\it JINR Communication E2-82-779}, Dubna, JINR, 1982.

\bibitem{24bogoslovsky1983} Bogoslovsky~G.Yu., The relativistic inert mass tensor, {\it Vestn. Mosk. Univ. Ser. Fiz. Astron.} {\bf 24} (1983), no.~1, 70--71.

\bibitem{25bogoslovsky1983} Bogoslovsky~G.Yu., On the local anisotropy of space-time, inertia and force f\/ields, {\it Nuovo Cimento B} {\bf 77} (1983), 181--190.

\bibitem{26bogoslovsky1983} Bogoslovsky~G.Yu., A generalized Klein--Gordon equation and Mach's principle, {\it Vestn. Mosk. Univ. Ser. Fiz. Astron.} {\bf 24} (1983), no.~3, 59--61.

\bibitem{27bogoslovsky&goenner2001} Bogoslovsky~G.Yu., Goenner~H.F., On the generalization of the fundamental f\/ield equations for locally anisotropic space-time, in Proceedings of XXIV International Workshop ``Fundamental Problems of High Energy Physics and Field Theory" (June 27--29, 2001, Protvino, Russia),
Editor V.A.~Petrov, Protvino, Insitute for High Energy Physics,
2001, 113--125,
\url{http://dbserv.ihep.su/~pubs/tconf01/c2-5.htm}.

\bibitem{28berwald1947} Berwald~L., Projective Kr\"ummung allgemeiner af\/f\/iner R\"aume und Finslersche R\"aume skalarer Kr\"ummung, {\it Ann.~Math.} {\bf 48} (1947), 755--781 (und die Literaturhinweise darin).

\bibitem{29moor1954} Mo\'or~A., Erg\"anzung, {\it Acta.~Math.} {\bf 91} (1954), 187--188.

\bibitem{30weyl191819} Weyl~H., Gravitation und Elektrizit\"at, {\it Sitzber. preuss Akad. Wiss., Physik--math. Kl.}, 1918, 465--480.\\
 Weyl~H.,    Eine neue Erweiterung der Relativit\"atstheorie, {\it Ann. Phys.} {\bf 59} (1919), 101--133.

\bibitem{31bogoslovsky1992} Bogoslovsky~G.Yu., From the Weyl theory to a theory of locally anisotropic space-time, {\it Classical Quantum Gravity} {\bf 9} (1992), 569--575.

\bibitem{32bogoslovsky2007} Bogoslovsky~G.Yu., 4-momentum of a particle and the mass shell equation in the entirely anisotropic space-time, in  Space-Time Structure (Algebra and Geometry), Editors D.G.~Pavlov, Gh.~Atanasiu  and   V.~Balan, Moscow, Lilia Print, 2007, 156--173.

\bibitem{33arbuzov2007} Arbuzov~B.A., Infrared non-perturbative QCD running coupling from Bogolubov approach, {\it Phys.~Lett.~B} {\bf 656} (2007), 67--73, \href{http://arxiv.org/abs/hep-ph/0703237}{hep-ph/0703237}.


\end{thebibliography}
\end{document}